\documentclass[aps,pra,twocolumn,notitlepage,showkeys,nofootinbib,showpacs,superscriptaddress,10pt,amssymb,amsmath]{revtex4-1}
\usepackage{graphicx}
\usepackage{calc}
\usepackage[pdftex]{xcolor}
\definecolor{darkblue}{rgb}{0.1921,0.2117,0.549}
\usepackage[pdftex,
        colorlinks=true,
        linkcolor=darkblue,
        citecolor=darkblue,
        filecolor=black,
        urlcolor=darkblue,
        bookmarks=true,
        bookmarksopen=true,
        bookmarksopenlevel=3,
        plainpages=false,
        pdfpagelabels=true,
        verbose=true]{hyperref}      
\newcommand{\tcd}[2][darkblue]{\textcolor{#1}{#2}}

\begin{document}
\title{Breaking inversion symmetry in a state-dependent honeycomb lattice: \\ Artificial graphene with tunable band gap}

\author{M.~Weinberg}
\affiliation{\mbox{Institut f\"{u}r Laserphysik, Universit\"{a}t Hamburg, Luruper Chaussee 149, D-22761 Hamburg, Germany}}
\author{C.~Staarmann}
\affiliation{\mbox{Zentrum f\"ur  Optische Quantentechnologien, Universit\"at Hamburg, Luruper Chaussee 149, D-22761 Hamburg, Germany}}
\author{C.~\"{O}lschl\"{a}ger}
\affiliation{\mbox{Institut f\"{u}r Laserphysik, Universit\"{a}t Hamburg, Luruper Chaussee 149, D-22761 Hamburg, Germany}}
\author{J.~Simonet}
\affiliation{\mbox{Institut f\"{u}r Laserphysik, Universit\"{a}t Hamburg, Luruper Chaussee 149, D-22761 Hamburg, Germany}}
\affiliation{\mbox{Zentrum f\"ur  Optische Quantentechnologien, Universit\"at Hamburg, Luruper Chaussee 149, D-22761 Hamburg, Germany}}
\author{K.~Sengstock}
\email{sengstock@physnet.uni-hamburg.de}
\affiliation{\mbox{Institut f\"{u}r Laserphysik, Universit\"{a}t Hamburg, Luruper Chaussee 149, D-22761 Hamburg, Germany}}
\affiliation{\mbox{Zentrum f\"ur  Optische Quantentechnologien, Universit\"at Hamburg, Luruper Chaussee 149, D-22761 Hamburg, Germany}}

\begin{abstract}
    Here, we present the application of a novel method for controlling the geometry of a state-dependent honeycomb lattice: The energy offset between the two sublattices of the honeycomb structure can be adjusted by rotating the atomic quantization axis. This enables us to continuously tune between a homogeneous graphene-like honeycomb lattice and a triangular lattice and to open an energy gap at the characteristic Dirac points. We probe the symmetry of the lattice with microwave spectroscopy techniques and investigate the behavior of atoms excited to the second energy band. We find a striking influence of the energy gap at the Dirac cones onto the lifetimes of atoms in the excited band.
\end{abstract}
\maketitle

The two dimensional honeycomb structure of graphene gives rise to an unusual electronic spectrum of massless, ultra-relativistic electrons as the two lowest energy bands touch at the vertices of the hexagonal Brillouin zone such that the dispersion relation becomes linear. For low energies this results in the formation of new quasiparticles that are accurately described by the three-dimensional Dirac equation \cite{Novoselov2005}. The touching points of the bands, the so-called Dirac points, exhibit a topological singularity with a localized Berry flux of $\pi$. Together with its extremely strong bonding and two-dimensional nature, the extraordinary topology of graphene gives rise to a large variety of fascinating phenomena, such as the anomalous quantum Hall effect and exceptional charge, heat and spin transport characteristics, only to name a few \cite{Geim2007,Geim2009}.

Ultracold atoms in unconventional optical lattices provide an ideal testing ground for such intriguing effects while also providing the possibility to engineer new systems with similar but complementary properties such as \textit{artificial graphene} \cite{Soltan-Panahi:2011ey,Soltan-Panahi:2011,Tarruell:2012db,Uehlinger2013,Polini2013,Jotzu2014,Duca2015,Messer2015,Li2015,Flaeschner2015}. These experiments allow for a high degree of control over internal and external system parameters, namely interactions, filling factors, tunneling properties to neighboring and next-neighboring lattice sites and particle species (bosonic or fermionic). In particular, the Dirac cones in \textit{artificial graphene} can be opened and closed dynamically to study topologically non-trivial band structures as well as higher-band dynamics as reported here.

The first experimental demonstration of \textit{artificial graphene} with quantum gas systems in 2011 \cite{Soltan-Panahi:2011ey,Soltan-Panahi:2011} studied bosonic spin mixtures and the Mott-insulator to superfluid transition. In addition, fermionic quantum gases have been used to study the merging of Dirac cones in a \textit{brick} lattice \cite{Tarruell:2012db}. Recently, the Haldane model was engineered \cite{Jotzu2014} and the singular Berry flux \cite{Duca2015}, Wilson lines \cite{Li2015} as well as Berry curvature \cite{Flaeschner2015} could be measured in an optical honeycomb lattice.

Recent years have seen an enormous increase of interest in the study of multi-orbital quantum gas systems. So far, experiments with bosonic quantum gases in optical lattice have successfully achieved the addressing of excited bands where Bose-Einstein condensation at nonzero quasimomenta could be observed in a three-dimensional cubic lattice \cite{Mueller2007} as well as in a bipartite chequerboard lattice \cite{Wirth:ohKKvtz7,Oelschlaeger:2011da}. In the latter case, evidence for a complex valued chiral order parameter was found and a topologically avoided band transition could be investigated \cite{Oelschlaeger2013a,Oelschlaeger:2012fw}. Higher energy bands could also be populated in fermionic systems by interaction-induced transitions between Bloch bands \cite{Koehl:2005cf} and momentum-resolved modulation spectroscopy \cite{Heinze:2011hf} that allowed the investigation of particle-hole dynamics \cite{Heinze2013}. Despite these achievements, corresponding investigations of decay channels are sparse and inter-band relaxation processes still lack a deeper understanding in quantum gas systems.

In this paper we present a graphene-like honeycomb lattice with a state-dependent component that allows us to continuously tune the symmetry of the lattice structure. The change of the lattice symmetry is accomplished by adjusting the energy offset between the diatomic basis of the graphene lattice. As one of the two sublattice sites becomes energetically favorable the inversion symmetry of the lattice is broken and changed from a homogeneous honeycomb lattice to a triangular one. With this, an energy gap can be created and tuned at the aforementioned Dirac cones.

We report on experiments probing the state-dependent \textit{artificial graphene} lattice with respect to the opening of the band gap at the Dirac point \cite{Luehmann2014hex}. In section \ref{sec1}\ref{sec1}, we explain the method used to control the lattice symmetry and, thus, the band structure by rotating the magnetic quantization field. The effect of the rotation on the lattice structure is characterized by performing band-selective microwave spectroscopy as detailed in section \ref{sec2}. This spectroscopy technique allows exciting the atomic ensemble to the second lowest energy band. The dynamic behavior of the atomic ensemble with respect to the presence of an energy gap at the Dirac cones is studied in section \ref{sec3}. Time-resolved measurements of the population of the excited band reveal a striking influence of the presence of a Dirac point. Section \ref{sec4} provides a conclusion of the presented results and an outlook to further possibilities for the study of ultracold quantum gases in graphene-like lattices.

\begin{figure}\hypertarget{fig:01ht}{}
    \centering
        \includegraphics{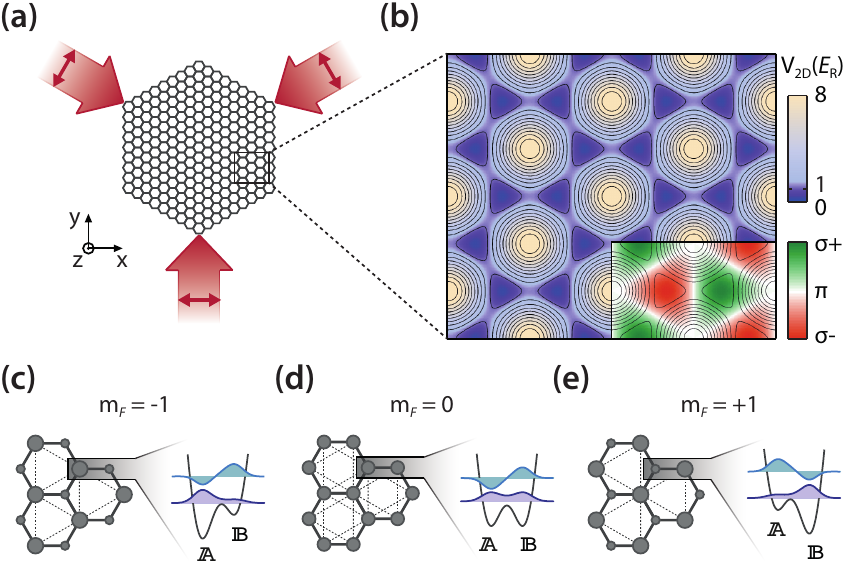}
        \caption[]{Spin-dependent honeycomb potential (a) Illustration of the three-beam setup for the generation of the spin-dependent honeycomb potential (not to scale). All three beams are linearly polarized along the lattice plane. (b) Resulting potential for an atom in the $m_F=0$ state with the additional alternating circular polarization pattern. (c) to (e) resulting lattice geometries for atoms in the spin-dependent lattice with $m_F=0,\pm1$. Atoms with non-vanishing magnetic quantum number are localized in one of the two triangular sublattices denoted $\mathbb{A}$ and $\mathbb{B}$.}
        \label{fig:01}
\end{figure}

\section{\label{sec1}State-dependent honeycomb lattice}
The state-dependent honeycomb lattice is generated by three running-wave laser beams that intersect in the $xy$-plane under angles of $120^\circ$ with an in-plane alignment of the linear polarization vectors $\boldsymbol{\epsilon}_i$ of the three-beam lattice setup \cite{Grynberg:1993bm,Becker:2010de,Luehmann2014hex} as illustrated in \hyperlink{fig:01ht}{\ref{fig:01}(a)}. This orientation of the polarization vectors generally gives rise to an alternating pattern of circular polarization in the resulting light field [see Fig.\,\hyperlink{fig:01ht}{\ref{fig:01}(b)}] which stems from the projection of the oscillating electric field onto the quantization axis, defined by a homogeneous magnetic field. Considering $^{87}\text{Rb}$ atoms and a laser wavelength of $\lambda_L=830\,\text{nm}$, the lattice laser detuning relative to the atomic transitions still is of the order of the fine structure splitting. In addition to the intensity modulation $V_\text{int}(\mathbf{x})$ of the resulting light field, this circumstance gives rise to a reasonably strong polarization-induced Stark shift of the magnetic Zeeman sub-states $|F,m_F\rangle$.  The total optical potential can be expressed as a sum of a state-independent and a state-dependent part:
\begin{equation}
	V(\mathbf{x})= -\frac{V_0}{8}\Big[V_{\text{int}}(\mathbf{x}) + V_{\text{pol}}(\mathbf{x})\Big].
\end{equation}
Here, $V_0$ denotes the corresponding lattice depth created for two equivalent counter-propagating laser beams. It is commonly given in units of the recoil energy $E_R=h^2/(2M\lambda_L^2)$, where $h$ is the Planck constant and $M$ the atomic mass of $^{87}$Rb. The state-independent part of the potential reads
\begin{equation}\label{eq:IntPotential}
V_\text{int}(\mathbf{x}) = 6 - 2 \sum_i\cos(\mathbf{b}_i\mathbf{x}),
\end{equation}
where each two reciprocal lattice vectors $\mathbf{b}_i = \varepsilon_{ijk}(\mathbf{k}_j - \mathbf{k}_k)$, given by the corresponding wave vectors of the laser beams $\mathbf{k}_1=2\pi(0,1,0)/\lambda_L$,  $\mathbf{k}_{2/3}=\pi(\pm\sqrt{3},-1,0)/\lambda_L$, span the reciprocal Bravais lattice. The state-dependent part of the optical potential can be obtained by calculating the projection of the light field onto the polarization basis vectors $\boldsymbol\varepsilon_{\mathcal{P}}$, with polarization $\mathcal{P}=\{\pi,\sigma^+,\sigma^-\}$. This basis is determined by the orientation of the systems quantization axis, which can be easily controlled in experiment by a homogenous magnetic field. In case of the quantization axis pointing along the $z$-axis, the $\boldsymbol\varepsilon_{\mathcal{P}}$ are the three-dimensional Jones vectors $\boldsymbol\varepsilon_\pi=(0,0,1)$ and $\boldsymbol\varepsilon_{\sigma^\pm}=(1,\pm i,0)/\smash{\sqrt{2}}$. For an arbitrary orientation of the quantization axis the basis has to be transformed in order to ensure that $\boldsymbol\varepsilon_\pi$ remains parallel to the quantization axis: $\boldsymbol\varepsilon_{\mathcal{P}}\rightarrow \boldsymbol\varepsilon_{\mathcal{P}}'= R_z(\gamma)R_x(\beta)R_y(\alpha)\boldsymbol\varepsilon_{\mathcal{P}}$. Here, the, $\alpha, \beta, \gamma$ denote the Euler angles, defining the orientation of the quantization axis and the $R_i$ are the Cartesian rotation matrices. Without loss of generality, we restrict the following considerations to cases $\beta=0$. If we also set $\gamma=0$, the resulting state-dependent part of the potential reads
\begin{equation}\label{eq:PolPotential}
    V_\text{pol}(\mathbf{x}) = \sqrt{3}(-1)^{F}m_F\eta \cos(\alpha) \sum_i \sin(\mathbf{b}_i\mathbf{x}),
\end{equation}
where the relative strength of the state-dependent potential is given by the proportionality factor $\eta$ that is determined by the detuning of the lattice light with respect to the atomic transitions.

As an important prerequisite for our studies the alternating pattern of circular polarizations breaks the inversion symmetry of the honeycomb lattice for magnetic quantum numbers different from zero: the potential energy of one of the two triangular sublattices $\mathbb{A}$ and $\mathbb{B}$ is lifted while the other one is lowered [compare Fig.\,\hyperlink{fig:01ht}{\ref{fig:01}(c)} to \hyperlink{fig:01ht}{(e)}]. Whether an atom in a hyperfine state $\left|F,m_F\right\rangle$ is predominantly confined at a lattice site with $\sigma^+$ or $\sigma^-$ polarization depends on the sign of its magnetic quantum number and the respective Land\'e factor $g_F$. Hereby, the prefactor $(-1)^F$ in Eq.\,(\ref{eq:PolPotential}) incorporates the sign change of the Land\'e factor in the ground-state manifold of $^{87}\text{Rb}$. Accordingly, an atom in the hyperfine state $\left|1,-1\right\rangle$ experiences the same potential as an atom in the state $\left|2,+1\right\rangle$ and both are mainly trapped in the same sublattice.

\begin{figure*}\hypertarget{fig:02ht}{}
    \centering
        \includegraphics{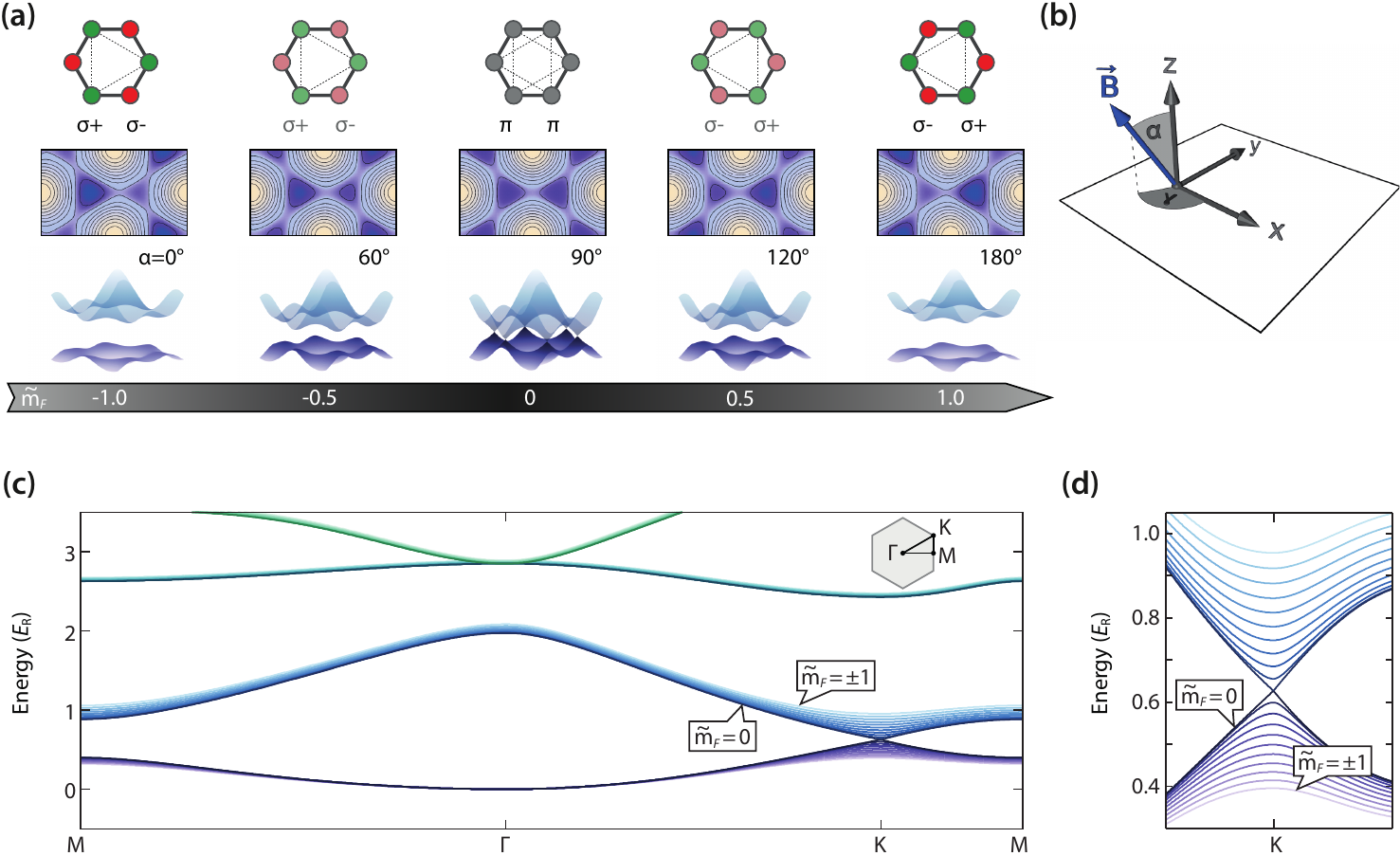}
        \caption[]{Effective magnetic quantum number and continuous opening of Dirac points. (a) A rotation by the Euler angle $\alpha$ continuously changes the effective magnetic quantum number $\tilde{m}$ allowing for the precise adjustment of the circular polarization component of the light field and, thus, the energy offset between sublattice sites. (b) Illustration of the quantization field in the defined coordinate system with two Euler angles $\alpha$ and $\gamma$. The three-beam lattice is aligned in the \textit{xy}-plane. Parts (c) and (d) depict the continuous opening of the Dirac point at the K-point of the Brillouin zone for a lattice depth of $1\,E_\mathrm{R}$. Here, the different lines correspond to values of $\tilde{m}$ ranging from $0$ to $\pm1$ in steps of $0.1$.}
        \label{fig:02}
\end{figure*}

Due to its linear dependency on the magnetic quantum number $m_F$ (resembling the linear Zeeman effect), the state-dependent part of the potential can be expressed as an effective magnetic field:
\begin{align}
    V_{\text{2D}}(\mathbf{r}) = V_{\text{Int}}(\mathbf{r}) + g_F\tilde{m}\mu_{\mathrm{B}}B_{\text{eff}}(\textbf{r}).\label{eq6beff}
\end{align}
where $\tilde{m}$ is an \textit{effective} magnetic quantum number that incorporates the magnetic quantum number, the respective Land\'e-factor and the rotation of the quantization field:
\begin{align}
    \tilde{m} = (-1)^Fm_F\cos(\alpha).
\end{align}
The strength of the energy offset between the twofold basis of the honeycomb lattice can thus be continuously tuned for a single hyperfine state by rotating the quantization field by the Euler angle $\alpha$ around the $y$-axis [see Fig.\,\hyperlink{fig:02ht}{\ref{fig:02}(b)}].

\begin{figure}[t]\hypertarget{fig:03ht}{}
    \centering
        \includegraphics{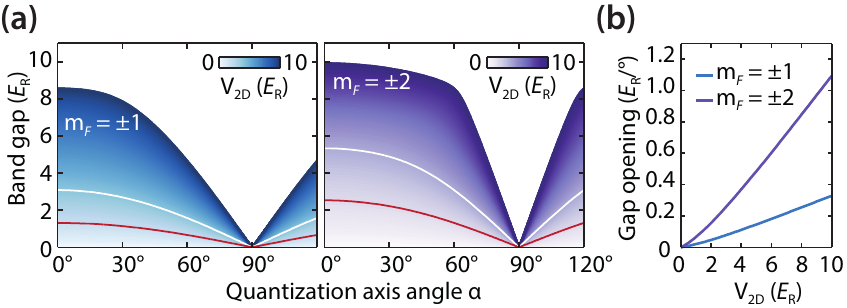}
        \caption[]{Opening of the band gap for increasing lattice depths. (a) The band gap at the K-point is plotted for increasing lattice depths with respect to the Euler angle $\alpha$. Red and white solid lines indicate a lattice depth of $2\,E_{\mathrm{R}}$ and $4\,E_{\mathrm{R}}$ respectively. (b) The linear slope of the band gap opening close to $\alpha=90^\circ$ is plotted in dependence of the lattice depth in units of $E_{\mathrm{R}}$ per degree.}
        \label{fig:03}
\end{figure}

The effect of the rotation of the quantization field onto the total lattice potential and its band structure is shown in Fig.\,\ref{fig:02} for the case of the $^{87}\mathrm{Rb}$ hyperfine states $\left|1,+1\right\rangle$ or $\left|2,-1\right\rangle$. As depicted in Fig.\,\hyperlink{fig:02ht}{\ref{fig:02}(a)}, the potential energy for these states is always lowered for $\sigma^+$. By rotating the quantization field $\mathbf{B}(\mathbf{r})$ away from its initial orientation parallel to the $z$-axis for an angle of $\alpha = 0^\circ$ the strength of the circular polarization components in the light field decreases. If the quantization field is exactly aligned in the $xy$-plane of the lattice, i.e. $\alpha = 90^\circ$, the polarization of the light field is everywhere $\pi$. Now the two sublattices are energetically degenerate such that the perfectly symmetric honeycomb lattice is restored, corresponding to $\tilde{m}=0$, and the two lowest energy bands touch at the vertices of the first Brillouin zone forming Dirac cones. Rotating the quantization field further re-opens the energy gap while the circular polarization pattern is interchanged: Sites with previously $\sigma^+$ polarization exhibit a $\sigma^-$ polarization and vice versa.

The corresponding band structure and opening of the Dirac points in dependence of the effective magnetic quantum number is plotted in Fig.\,\hyperlink{fig:02ht}{\ref{fig:02}(c)} and \hyperlink{fig:01ht}{(d)} for a lattice depth of $V_{\text{2D},0} = 1\,E_{\text{R}}$. While the energy gap is perfectly closed for an effective quantum number of $\tilde{m}=0$ it already amounts to $E_{\mathrm{gap}}\approx0.6\,E_{\mathrm{R}}$ for $\tilde{m}=\pm1$ which corresponds to approximately $2\,\mathrm{kHz}$.

The quantitative behavior of the energy gap opening with respect to the Euler angle $\alpha$ is depicted in Fig.\,\ref{fig:03}. In part \hyperlink{fig:02ht}{\ref{fig:03}(a)} the calculated band gap at the K-points of the Brillouin zone are plotted for increasing lattice depths up to $10\,E_{\mathrm{R}}$ in dependence of $\alpha$ for initial magnetic quantum numbers of $m_F=\pm1$ and $m_F=\pm2$. Hereby, white and red lines indicate a lattice depth of $2\,E_{\mathrm{R}}$ and $4\,E_{\mathrm{R}}$ respectively. The band gap increases linearly close to angles of $\alpha=90^\circ$ for all lattice depths. In Fig.\,\hyperlink{fig:03ht}{\ref{fig:03}(b)}, the corresponding slope of the band opening is plotted for both initial magnetic quantum numbers in units of $E_{\mathrm{R}}$ per degree of $\alpha$ for increasing lattice depths. For an accurate control over the opening and closing of the Dirac cone it is, thus, crucial to be able to precisely adjust the Euler angle $\alpha$. In the experimental setup the alignment of the quantization field is achieved by a set of Helmholtz coils. Residual stray fields can be compensated fairly well up to the order of $\mathrm{mG}$ with additional sets of Helmholtz coils. As can be deduced from Fig.\,\hyperlink{fig:03ht}{\ref{fig:03}(b)}, a stray field of $1\,\mathrm{mG}$ perpendicular to the lattice plane corresponds to a band gap of approximately $20\,\mathrm{mHz}/E_{\text{R}}$ for a hyperfine state with $m_F=\pm1$ at a quantization angle of $\alpha=90^\circ$.

In conclusion, using different spin states of quantum gases in the state dependent honeycomb lattice allows to engineer a new form \textit{artificial graphene} in which the energy offset of the different sublattices $\mathbb{A}$ and $\mathbb{B}$ can be tuned via the rotation of an external quantization field such that the Dirac cones can be opened and closed. Note that, in addition, it is also possible to \textit{shift} the \textit{position} of Dirac points in momentum space by inducing anisotropic tunneling rates along different lattice directions, e.g., by an anisotropic intensity distribution in the three lattice laser beams as in Ref.\,\cite{Duca2015}.

\section{\label{sec2}Characterization of the lattice geometry via microwave spectroscopy}
The influence of the orientation of the quantization axis onto the lattice geometry can be elegantly probed by a band-selective microwave spectroscopy. This versatile technique was previously applied to investigate interaction effects and the spatial ordering of spin-mixtures in the state-dependent honeycomb lattice \cite{Soltan-Panahi:2011ey}. In the following, we briefly describe the basic principles and experimental procedure of band-selective microwave spectroscopy and apply the method to investigate the tunability of the effective magnetic quantum number $\tilde{m}$.

\begin{figure}[t]\hypertarget{fig:04ht}{}
    \centering
        \includegraphics{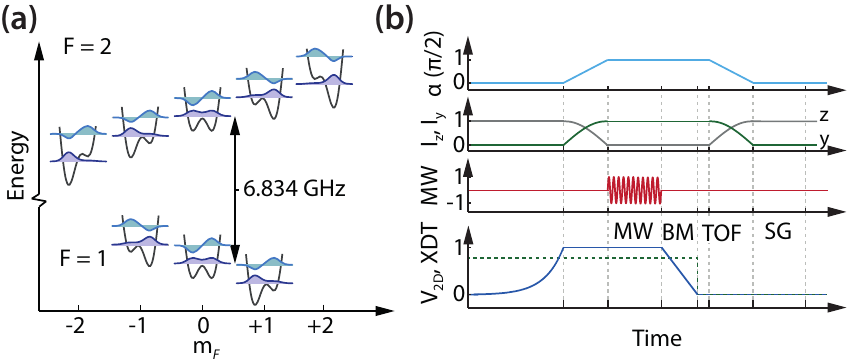}
        \caption[]{Level structure and experimental procedure. (a) The energy level structure of the two ground-state hyperfine manifolds $F=1,\,2$ is shown for the twofold atomic basis in the state-dependent honeycomb lattice. The two lowest Bloch bands are plotted for each magnetic sub-state. (b) Experimental procedure of the microwave spectroscopy and rotation of the magnetic field (not to scale). After the lattice potential $V_{\mathrm{2D}}$ (dark blue line) is ramped to its final intensity in $100\,\mathrm{ms}$ the quantization axis angle $\alpha$ (light blue) is rotated in approximately $0.5\,\mathrm{ms}$ by changing the currents $I_y$ (green) and $I_z$ (gray) through the Helmholtz coils. Subsequently, a microwave pulse (red) is applied to the atomic ensemble. Before the dipole trap (dashed green line) is shut off the lattice is completely ramped down in $1\,\mathrm{ms}$, transferring the quasimomentum distribution into real momenta (band mapping). After the crossed dipole trap (XDT, dashed green line) is switched off the quantization axis is rotated back to its initial orientation during the time-of-flight (TOF) before a Stern-Gerlach gradient field (SG) is applied for $12\,\mathrm{ms}$ that separates the spin components.}
        \label{fig:04}
\end{figure}

The experimental procedure of lattice characterization is illustrated in Fig.\,\hyperlink{fig:04ht}{\ref{fig:04}(b)}. First, the atomic ensemble in an initial hyperfine state $\left|F,m_F\right\rangle$ is adiabatically loaded into the spin-dependent optical lattice with an additional perpendicular 1D-lattice with a potential depth of $V_{\text{1D},0}$. After rotating the quantization axis by an angle $\alpha$, a microwave pulse is applied, which drives transitions between the two ground-state hyperfine manifolds $F=1,2$ of $^{87}\mathrm{Rb}$ [see Fig.\,\hyperlink{fig:04ht}{\ref{fig:04}(a)}]. The different spin components are detected separately after time-of-flight.

In the superfluid regime, the atomic ensemble occupies the lowest Bloch band of the lattice at zero quasimomentum $\psi_{\mathbf{q}=0}^0(\mathbf{r})$. In the following, we use the notation $\psi_{{s}}^n(\mathbf{r})$ for the $n$-th Bloch wave function at zero quasimomentum in the hyperfine state $s\equiv\left|F,m_F\right\rangle$. From the initial state in the zeroth Bloch band $\psi_{s}^0(\mathbf{r})$ transitions can be driven to other Bloch bands $\psi_{s'}^n(\mathbf{r})$. The microwave spectroscopy can be treated as a two-level problem such that transitions are described as a Rabi oscillation between the initial state $\psi_{s}^0(\mathbf{r})$ and the final state $\psi_{s'}^n(\mathbf{r})$. The ratio of transferred atoms is then given by
\begin{align}
    \frac{N'(t)}{N_0} = \frac{\Omega^2}{\Omega^2+\Delta\omega^2}\cdot\sin^2\left( \sqrt{\Omega^2+\Delta\omega^2}\cdot t/2   \right)\label{eq6sinc}
\end{align}
where the frequency detuning $\Delta\omega=\omega_{\text{MW}}-\Delta\omega_{\psi'\psi}$ is determined by the microwave frequency $\omega_{\text{MW}}$ and the frequency difference $\Delta\omega_{\psi'\psi}$ between the initial and the final hyperfine state in its respective Bloch band. $\Omega$ denotes the Rabi-frequency of the transition that relates to the on-resonance Rabi-frequency of free atoms $\Omega_R$ according to
\begin{align}
    \Omega = \Omega_R \int {\psi_{s'}^n}^{\ast}(\mathbf{r})\hspace{1mm} \psi_{s}^0(\mathbf{r})\hspace{1mm}\mathrm{d}^3r.
\end{align}
The spatial integral in the above equation defining the overlap between initial and final wave function is known as the Franck-Condon factor.

\begin{figure}[t]\hypertarget{fig:10ht}{}
    \centering
        \includegraphics{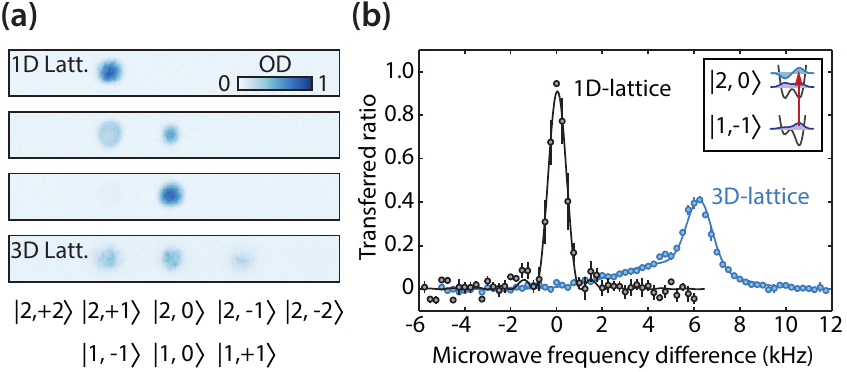}
        \caption[]{Microwave spectroscopy and spin dynamics. (a) Time-of-flight images of magnetic sub-states separated by a Stern-Gerlach field. The efficient transfer from the initial $|1,-1\rangle$ state into the $|2,0\rangle$ state is demonstrated in the one-dimensional lattice (first three images). (b) Microwave spectra for the transition obtained in the 1D and the 3D-lattice. While the transition in the state-independent 1D-lattice is exactly at the same position as for the free dipole trap (zero position of the microwave frequency difference) the transition in the state-dependent potential is clearly shifted towards larger frequencies. Error bars indicate the standard deviation of the data that were averaged at least twice. Solid lines depict fits to the spectra according to Eq.\,(\ref{eq6sinc}) for the 1D-lattice (black) and a double Gaussian for the 3D-lattice case (blue).}
        \label{fig:10}
\end{figure}

An example for a microwave transition from the initial $|1,-1\rangle$ state into the $|2,0\rangle$ state for $\alpha=0^{\circ}$ is shown in Fig.\,\ref{fig:10}. Here, the involved spin components are separated during time-of-flight by a Stern-Gerlach gradient field and detected by resonant absorption imaging [see Fig.\,\hyperlink{fig:10ht}{\ref{fig:10}(a)}]. The resulting transition spectra, depicted in Fig.\,\hyperlink{fig:10ht}{\ref{fig:10}(b)}, exhibit clear resonance features, whereby a clear shift of the transition frequency can be observed in the spin-dependent honeycomb lattice.

\begin{figure}\hypertarget{fig:05ht}{}
    \centering
        \includegraphics{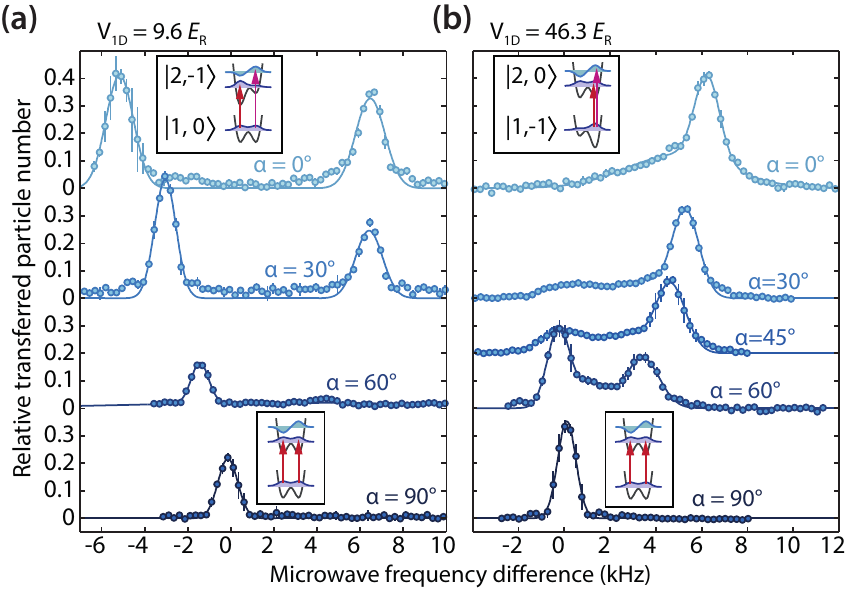}
        \caption[]{Microwave spectra of the tunable honeycomb lattice. (a) The transition from $\left|1,0\right\rangle$ to $\left|2,-1\right\rangle$ is shown for a honeycomb lattice depth of $V_{\text{2D},0} = 4.0\,E_{\text{R}}$ and an additional 1D-lattice of $V_{\text{1D},0} = 9.6\,E_{\text{R}}$. A shift of the transition between lowest Bloch state towards smaller frequencies can clearly be observed for increasing energy offset between the final two-fold atomic basis. In addition, a second transition into the first excited Bloch state appears for a strong energy difference (see red arrows in the inset). (b) A transition from $\left|1,-1\right\rangle$ to $\left|2,0\right\rangle$ is shown for the same honeycomb lattice depth but a strong 1D-lattice of $V_{\text{1D},0} = 46.3\,E_{\text{R}}$. Again, a transition to the first excited Bloch state emerges for increasing energy offset while a residual transition to the lowest Bloch state remains. Solid lines in (a) and (b) are Gaussian fits to the data.}
        \label{fig:05}
\end{figure}

In Fig.\,\ref{fig:05} two different scenarios are investigated with respect to the orientation of the quantization axis. The microwave spectra presented in Fig.\,\hyperlink{fig:05ht}{\ref{fig:05}(a)} are obtained for an initial state with magnetic quantum number $m_F=0$ that is independent of any circular polarization of the lattice light field while the final state with $m_{F}=-1$ is strongly influenced by the rotation of the quantization field. Hereby, the honeycomb lattice depth amounts to $V_{\text{2D},0} = 4.0\,E_{\text{R}}$ and the additionally confining 1D-lattice depth is $V_{\text{1D},0} = 9.6\,E_{\text{R}}$. In contrast, the spectra shown in Fig.\,\hyperlink{fig:05ht}{\ref{fig:05}(b)} are obtained for the opposite case of a state-independent final hyperfine state and a state-dependent initial state. While the honeycomb lattice depth is also set to  $V_{\text{2D},0} = 4.0\,E_{\text{R}}$, the 1D-lattice potential of  $V_{\text{1D},0} = 46.3\,E_{\text{R}}$ significantly increases on-site interaction effects.

For $\alpha = 90^\circ$, the quantization axis is aligned exactly in the lattice plane and the twofold atomic basis of the honeycomb lattice is completely degenerate. Both spectroscopy signals are thus identical. Indicated by the respective insets, both the initial and final state have an effective magnetic quantum number $\tilde{m}=0$. Here, a single resonance is observed that corresponds to a transition into the lowest Bloch function of the final state. The transition into the second Bloch state is not observed as the Franck-Condon overlap between the initial and final wave functions vanishes: Both final and initial state are eigenstates of the isotropic honeycomb potential and, thus, orthonormal. This circumstance, however, significantly changes for an effective magnetic quantum number different from zero.

Let us first consider the transition $\left|1,0\right\rangle\rightarrow\left|2,-1\right\rangle$ in Fig.\,\hyperlink{fig:05ht}{\ref{fig:05}(a)}. If the quantization axis is rotated out of the lattice plane the resonance position for the transition between lowest bands first gets shifted towards smaller frequencies as evident for the case of $\alpha = 60^\circ$. This behavior can be understood as the final Bloch state $\psi_{\left|1,-1\right\rangle}^0$ becomes more deeply trapped at the now stronger confining sublattice site. Hence, the energy difference between initial and final state decreases due to the light shift. For larger energy offsets between the twofold atomic basis the resonance is shifted further towards smaller frequencies as observed in the spectra for $\alpha = 30^\circ$ and $60^\circ$. In addition, a second resonance appears that indicates a transition into the second Bloch state. This transition emerges since the initial and final potentials now strongly differ such that the respective Bloch functions of initial and final state are no longer orthonormal. Here, the change from an asymmetric triangular state to a symmetric honeycomb state is evident as the energy offset between the sublattice sites decreases.

\begin{figure}\hypertarget{fig:06ht}{}
    \centering
        \includegraphics{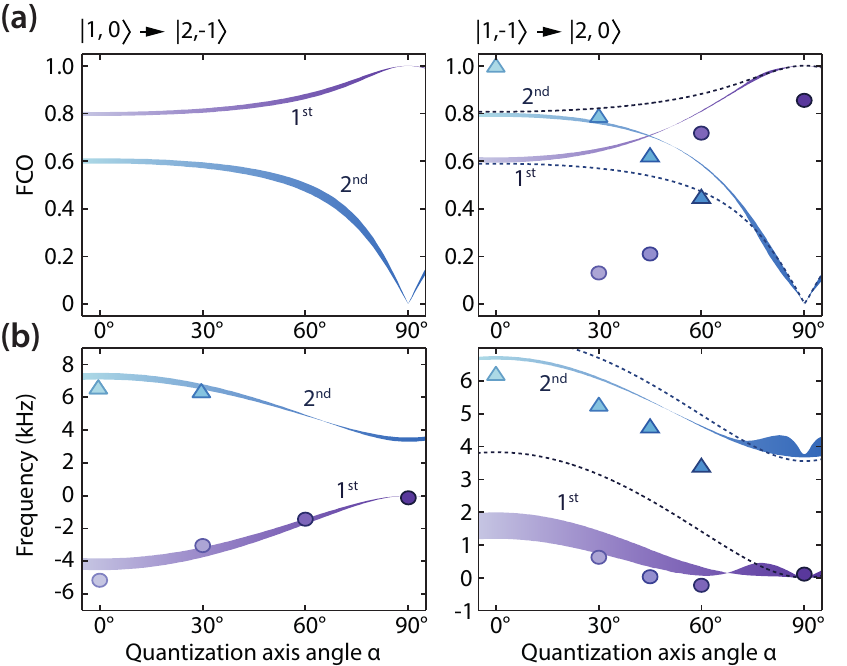}
        \caption[]{Quantitative investigation of microwave spectra. (a) Calculated Franck-Condon factors for the two investigated transitions (and the extracted values for the strongly interacting system) are plotted in dependence of $\alpha$. As expected the Franck-Condon overlap for the transitions to the second Bloch state vanishes for both cases in the symmetric honeycomb lattice at $\alpha =90^\circ$ where initial and final states are similar. (b) \textit{Ab initio} calculations of the resonance positions reproduce the extracted values (circles and triangles). While interaction effects are omitted for the calculations of the $\left|1,0\right\rangle\rightarrow\left|2,-1\right\rangle$ transition, the  $\left|1,-1\right\rangle\rightarrow\left|2,0\right\rangle$ transition is investigated by the iterative Gross-Pitaevskii method described in the text. Dashed lines indicate the calculated resonance positions neglecting interaction effects for the transition $\left|1,-1\right\rangle\rightarrow\left|2,0\right\rangle$. }
        \label{fig:06}
\end{figure}

The corresponding microwave spectra of the investigated $\left|1,-1\right\rangle\rightarrow\left|2,0\right\rangle$ transition are shown in Fig.\,\hyperlink{fig:05ht}{\ref{fig:05}(b)}. A similar emergence of a transition to the second Bloch state can be observed for an increasing energy offset. Moreover, this resonance shifts towards higher transition frequencies as now the energetically lower \textit{initial} state of the $F=1$ hyperfine manifold rather than the final state becomes increasingly confined such that the transitions are further separated. Strikingly, a similar shift of the transition to the first Bloch state is not reproduced. While the resonance position remains relatively constant for an increasing energy offset, it quickly becomes broader and vanishes. In contrast to the previously discussed  $\left|1,0\right\rangle\rightarrow\left|2,-1\right\rangle$ transition such a behavior cannot be understood in a simple single-particle picture.

In Fig.\,\ref{fig:06} the calculated Franck-Condon factors and the resonance positions of the first two Bloch states are shown for the transitions $\left|1,0\right\rangle\rightarrow\left|2,-1\right\rangle$ and $\left|1,-1\right\rangle\rightarrow\left|2,0\right\rangle$. In the non-interaction limit the theory predicts that the absolute values of both transition frequencies and Franck-Condon coincide (dashed lines). The finite spread of the calculated wave function overlap and the resonance positions indicate an assumed uncertainty of the calibrated lattice depths by $\pm5\%$. Our experimental results for the  $\left|1,-1\right\rangle\rightarrow\left|2,0\right\rangle$ transition are in strong contradiction to this prediction.

Indeed, for the $\left|1,-1\right\rangle\rightarrow\left|2,0\right\rangle$ transition, with a potential depth of $V_{\text{1D},0} = 46.3\,E_{\text{R}}$, the additional confining 1D-lattice was chosen to be significantly stronger as compared to the $\left|1,0\right\rangle\rightarrow\left|2,-1\right\rangle$ transition ($9.6\,E_{\text{R}}$). In addition, the initial state is confined to a single lattice site for quantization axis angles different from $\alpha=90^\circ$. This inhomogeneous density distribution strongly influences the final state, which experiences an effective potential deformed by the repulsive interparticle interactions. To gain a deeper understanding of the strength of this effect, the eigenfunction of $\left|2,0\right\rangle$ is numerically evaluated by solving coupled Gross-Pitevskii equations including the additional interaction shift. For the calculations, homogeneous average filling of $N=2.5$ to $3.5$ particles per lattice site was estimated while the calibrated lattice depths were assumed to be exact. This initial occupation number strongly affects the influence of the effective interaction potential giving rise to the areas of uncertainty in the depicted calculations. In comparison to calculations neglecting interaction effects (dashed lines), the obtained values for the Frank-Condon overlaps show a significant increase of the transition to the second band. Moreover, the observed shift of resonance frequencies for an increasing site-offset is confirmed by the iterative mean-field method in stark contrast to the non-interacting case. The observed resonance positions from the microwave spectra in Fig.\,\ref{fig:05}, indicated by circles (triangles) for the transition to the first (second) Bloch state are in good agreement with the \textit{ab initio} calculations.

In conclusion, the agreement of the investigated microwave spectra with the expected behavior of atomic ensembles in the state-dependent honeycomb lattice verifies the experimental technique of continuously tuning the band structure by altering the potential in real-space. Moreover, strong influences of interaction effects could be reproduced by means of an additional effective mean-field potential.

\section{\label{sec3}Probing inter-band dynamics}
While the interest in multi-orbital systems of ultracold atoms has increased tremendously in recent years, mechanisms responsible for inter-band relaxation remain elusive and such decay processes still lack a deeper understanding.

In contrast to solid state systems where a variety of mechanisms may induce band decay processes (e.g., charge carrier multiplication or exchange of energy with spin degrees of freedom) inter-band dynamics of ultracold atoms is driven by collisional processes \cite{Martikainen2011,Buecker2011,Paul2013}. For example, Heinze \textit{et al.} \cite{Heinze2013} report a striking influence of the scattering length on the lifetime of a particle-like excitation in the second energy band of single-component Fermi gas.

In the following, we aim to investigate decay mechanisms of bosonic atoms in excited bands of the tunable honeycomb lattice with respect to the presence of Dirac cones. By utilizing the possibility to continuously alter both the real-space lattice potential and, thus, the energy band structure the gap at the Dirac cones can be opened and closed in a controlled manner. The band-selective microwave spectroscopy, discussed in the previous section, is now used to transfer the atomic ensemble into excited Bloch bands in a controlled manner.

\begin{figure}\hypertarget{fig:07ht}{}
    \centering
        \includegraphics{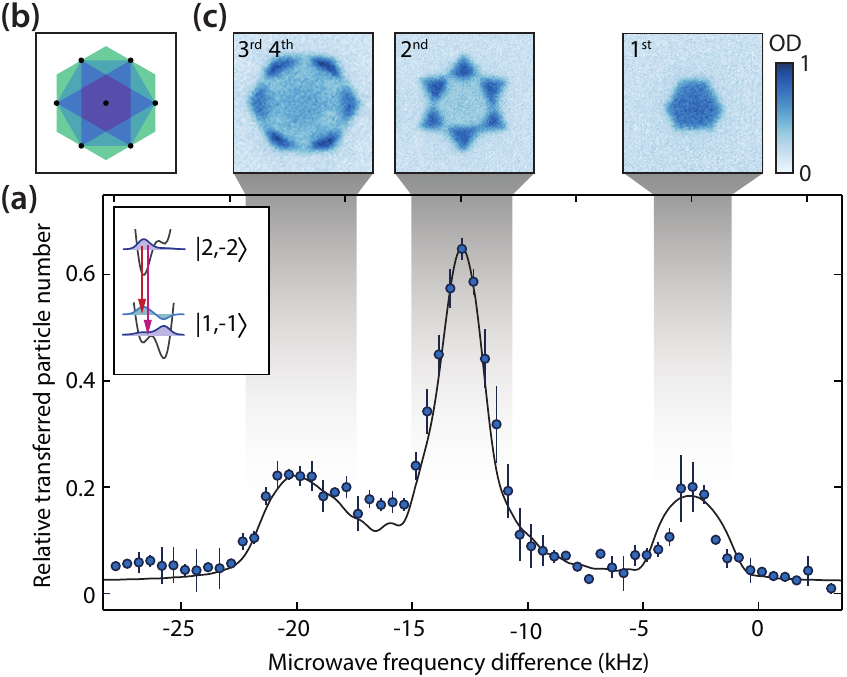}
        \caption[Addressing of higher bands with microwave transitions.]{Addressing of higher bands with microwave transitions. (a) The microwave spectrum of the transition from the $\left|2,-2\right\rangle$ to the $\left|1,-1\right\rangle$ state (see inset) exhibits three distinct resonances that can be attributed to the four lowest energy bands. Band mapping images (c) reveal the characteristic shape and population of the corresponding Brillouin zone (b), where the third and fourth band are simultaneously populated since they are degenerate at the $\Gamma$ point [compare Fig.\,\ref{fig:02}\tcd{(c)}]. The solid line indicates a theoretical calculation of the transition spectrum with the only free parameter being the absolute height of the transition.}
        \label{fig:07}
\end{figure}

In Fig.\,\hyperlink{fig:07ht}{\ref{fig:07}(a)}, a microwave spectrum of the transition from the  $\left|2,-2\right\rangle$ state into the $\left|1,-1\right\rangle$ state is shown for a quantization field aligned along the $z$-axis ($\alpha=0^\circ$). It is obtained at lattice depths of $V_{\text{2D},0} = 3.0\,E_{\text{R}}$ and $V_{\text{1D},0} = 4.0\,E_{\text{R}}$. By applying a band mapping procedure, information of the momentum-resolved band population can be obtained. Here, transitions to different Bloch bands can be identified [compare Fig.\,\hyperlink{fig:07ht}{\ref{fig:07}(b)} and \hyperlink{fig:07ht}{(c)}]. The resonance with the highest frequency is the transition to the lowest Bloch band where the corresponding band mapping image clearly reproduces the hexagonal shape of the first Brillouin zone. While this transition is relatively weak due to the small spatial overlap between the wave functions that are primarily confined to different lattice sites (see inset), the transition to the second Bloch state is strongly pronounced. Here, the band mapping image shows the characteristic hexagram of the second Brillouin zone. For smaller transition frequencies, a broad transition into both the third and the fourth energy band can be observed. These transitions occur simultaneously as they are degenerate at zero quasimomentum. Again, the band mapping clearly shows the population of the respective Brillouin zones. Note that, in contrast to the spectroscopy signal itself, the shown images were obtained by a \textit{sweep} of the microwave frequency over the resonance (indicated by gray shaded areas). With this, an optimal transfer into the given state can be achieved. The solid line in Fig.\,\hyperlink{fig:07ht}{\ref{fig:07}(a)} shows a theoretical calculation of the transition spectrum. Despite the relatively small lattice depths, mean-field interaction effects have been taken into account, whereby the initial on-site occupation number has been weighted according to the in-trap Thomas-Fermi density profile. In order to reproduce the shape of the transition, sinc-profiles determined by Eq.\,(\ref{eq6sinc}) have been summed for all four lowest transitions at every evaluated frequency. With the absolute height of the transition being the only free parameter in the numerical calculations the obtained signal is in excellent agreement with the experimental data.

Results shown in Fig.\,\ref{fig:07} demonstrate that sweeping the microwave pulse frequency through a transition to a final Bloch state is a suitable technique to transfer an atomic ensemble with high efficiency into excited Bloch bands of an optical lattice. In the following, we employ this method in order to investigate the dynamical behavior of atoms excited to the second Bloch band with respect to the presence of Dirac cones between the two lowest bands that can be adjusted by the rotation of the quantization axis.

\begin{figure}[t]\hypertarget{fig:08ht}{}
    \centering
        \includegraphics{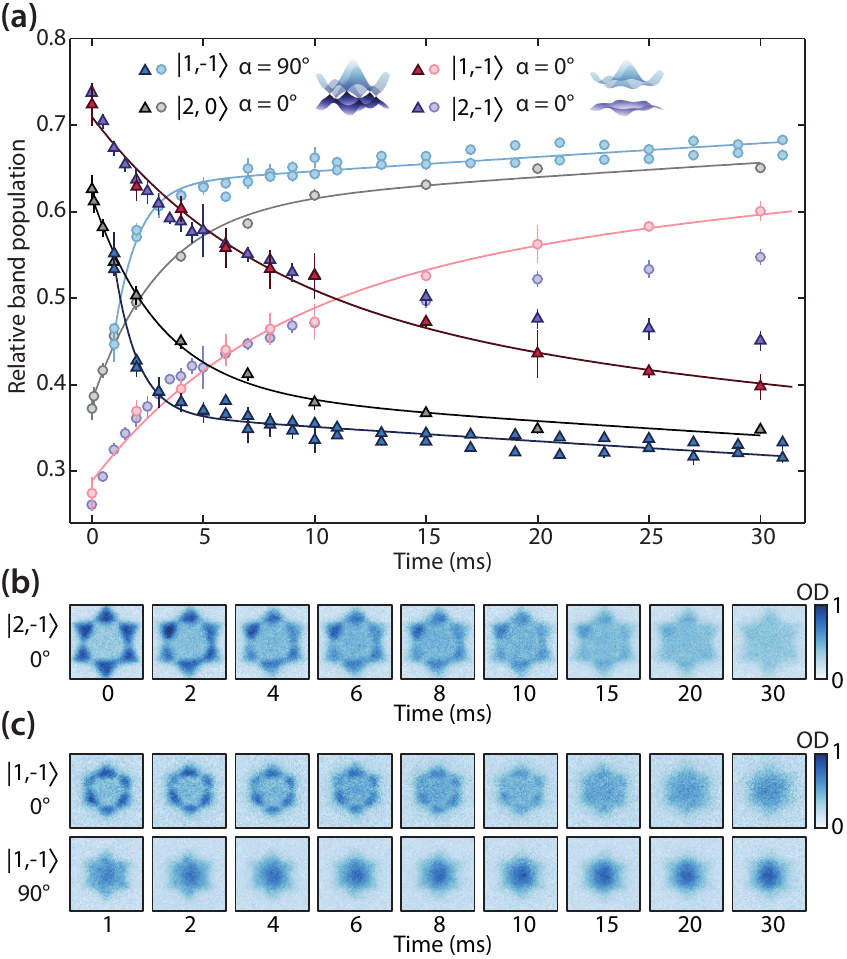}
        \caption[]{Band decay dynamics in the presence of Dirac cones. (a) The time-dependent relative population of the first (second) Brillouin zone is depicted by circles (triangles) for the two fundamentally different scenarios of a maximally opened band gap (red and purple markers) and closed Dirac points (blue and gray markers). Data is obtained by summing the atom numbers in the respective Brillouin zones. A striking difference of the lifetimes in the first excited Bloch band can be observed: While for both cases a finite fraction of atoms remains in the second Brillouin zone at all times, the initial decay to the first Brillouin zone is roughly an order of magnitude larger in the case of closed Dirac points. Error bars depict the standard deviation of multiple measurements. The two sets of blue markers stem from two independent measurement series. Solid lines represent double-exponential fits to the data. (b) Corresponding time-of-flight images of the observed band populations for the opened band gap and final state $\left|2,-1\right\rangle$. (c) Comparison of time-of-flight images of the final state $\left|1,-1\right\rangle$ for opened (closed) Dirac points at $\alpha=0^\circ$ ($\alpha = 90^\circ$).}
        \label{fig:08}
\end{figure}

In Fig.\,\hyperlink{fig:08ht}{\ref{fig:08}} time-dependent occupations of the first two Brillouin zones are shown for the two different scenarios: First, decay dynamics are investigated in a completely homogeneous honeycomb lattice, i.e., degenerate sublattices $\mathbb{A}$ and $\mathbb{B}$ for a vanishing effective quantum number $\tilde{m}=0$ where the Dirac cones are closed. Here, such a system is realized by both a final state of $|2,0\rangle$ at a quantization axis angle of $\alpha=0^\circ$ as well as a final state of $|1,-1\rangle$ at $\alpha=90^\circ$. Measurements were performed in the three-dimensional lattice with potential depths of $V_{\text{2D},0} = 3.0\,E_{\text{R}}$ and $V_{\text{1D},0} = 3.7\,E_{\text{R}}$. The final state is prepared in the first excited band by a microwave sweep (compare Fig.\,\ref{fig:07}) and band mapping is applied after a waiting time allowing to resolve the relative time-dependent population of the first and second Brillouin zone depicted in Fig.\,\hyperlink{fig:08ht}{\ref{fig:08}(a)} as circles and triangles respectively. Here, the time axis corresponds to the waiting time in the lattice system following the microwave sweep of $1\,\mathrm{ms}$ duration. Band populations are counted by summing over the area of the respective Brillouin zone after band mapping [see Fig.\,\hyperlink{fig:07ht}{\ref{fig:07}(a)}].

As a central result, for both the final states $|2,0\rangle$, $\alpha=0^\circ$ and the $|1,-1\rangle$, $\alpha=90^\circ$ a rapid decay of the population of the second Brillouin zone can be observed on the order of a few milliseconds. In a second set of experiments, in contrast to the closed Dirac points for $\tilde{m}=0$, the same time-dependent occupation is investigated with a \textit{maximally opened gap} between the lowest and first excited energy band for $\tilde{m}=\pm1$, i.e., for the final states of $|1,-1\rangle$ and $|2,-1\rangle$ at a quantization axis angle of $\alpha=0^\circ$ each. Here, a strikingly different behavior can be observed as the lifetime in the first excited band increases for more than an order of magnitude compared to the $\tilde{m}=0$ case. In the following we discuss this effect in more detail.

Examples of the corresponding band-mapping images obtained after time-of-flight are shown in Fig.\,\hyperlink{fig:08ht}{\ref{fig:08}(b)} for the final state $\left|2,-1\right\rangle$ at $\alpha=0^\circ$. In Fig.\,\hyperlink{fig:08ht}{\ref{fig:08}(c)} the difference of the decay timescales between the two scenarios is emphasized as time-of-flight images for the same initial state with opened and closed bands are shown. While in the case of a broken inversion symmetry, i.e., $\alpha=0^\circ$ and $\tilde{m}=+1$ the first Brillouin zone is populated by a majority of atoms after waiting times larger than $10\,\mathrm{ms}$, this occurs already after approximately $1.5\,\mathrm{ms}$ for the case of closed Dirac points. The corresponding time-dependent populations of the first two Brillouin zones are well reproduced by double-exponential fits to the data that are depicted in Fig.\,\hyperlink{fig:08ht}{\ref{fig:08}(a)} as solid lines.

The condition for the adiabaticity of the band mapping procedure cannot be fulfilled for an effective magnetic quantum number of $\tilde{m}=0$ as the energy gap between the investigated bands vanishes. Thus, in order to avoid an additionally induced transfer of atoms between the respective bands, we rotate the quantization axis back to its initial perpendicular orientation \textit{prior} to the band mapping process instead of performing this step during the time-of-flight. By this, the constraint $h/T_{\text{BM}} \ll E_{\text{gap}}$ imposed on the lattice ramping time $T_{\text{BM}}$ is fulfilled such that no additional inter-band transfers occur. The duration of the lattice rotation is reflected in the data as measurements start for a finite waiting time of $1\,\mathrm{ms}$ in this case (blue markers). For the same reason, the validity of observations corresponding to a final state of $\left|2,0\right\rangle$ (gray markers) is limited: The band gap cannot be opened in this case.

For bosonic atoms in higher bands of optical lattices, the decay mechanisms are driven by the interactions. Two collisional channels have been identified so far. A collision of atoms in the first excited Bloch band can result in both atoms decaying into the lowest energy band \cite{Paul2013} or one atom being excited into a higher band while the other atom decays into the lowest band \cite{Martikainen2011}, whereby both processes clearly are only allowed if energy and momentum can be conserved. In the former case the excess energy of the band gap is expected to be redistributed into an additional degree of freedom that can, e.g., be given by excitations along a weakly confining axis. However, experiments were performed in a true three-dimensional lattice with an additional perpendicular 1D-lattice confinement on the same order as the honeycomb lattice depth such that spatial excitations can be ruled out as the reservoir of excess energy.

\begin{figure}[t]\hypertarget{fig:09ht}{}
    \centering
        \includegraphics{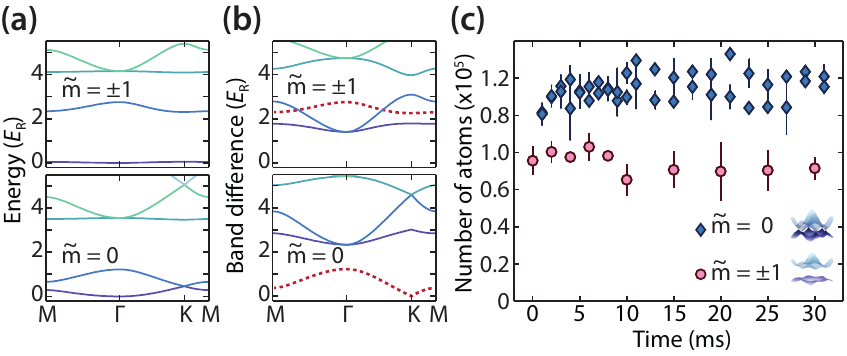}
        \caption[]{Possible inter-band collisional mechanism and total atom number for the two lowest energy bands. (a) Band structures of the investigated system in Fig.\,\ref{fig:08} are plotted for the cases of effective quantum numbers $\tilde{m}=\pm1$ and $\tilde{m}=0$. In stark contrast to the symmetric honeycomb lattice with intersecting lowest energy bands, a large energy gap between the first Bloch bands is opened for $\tilde{m}=\pm1$, i.e., $m_F=\pm1$ and $\alpha = 90^\circ$. (b) Energy differences between the second band and all other bands are plotted. The difference from the second to the first band is shown as a dashed red line. An intersection of transitions for $\tilde{m}=\pm1$ enables the decay channel of a simultaneous excitation to the fourth band. (c) Total atom number in the first two Brillouin zones for the measurements with a final state of $\left|1,-1\right\rangle$ reveal a constant behavior for both the opened and the closed band gap.}
        \label{fig:09}
\end{figure}

The scenario of exciting one atom into a higher Bloch band while another one decays into the lowest band described in Ref.\,\cite{Martikainen2011} is further investigated in Fig.\,\ref{fig:09}. The band structure of the state-dependent honeycomb lattice is shown in Fig.\,\hyperlink{fig:09ht}{\ref{fig:09}(a)} for the two distinct cases of an effective quantum number of $\tilde{m}=\pm1$ and $\tilde{m}=0$. The close-lying lowest bands for the symmetric lattice and the large opened band gap for a maximum energy offset between the sublattices are clearly visible. In part \hyperlink{fig:09ht}{(b)} of the figure, the corresponding energy differences between the second Bloch band and all other bands are plotted. Hereby, the difference between the second and the first band is emphasized as a dashed red line. Energy conservation of the discussed decay process requires the band difference to coincide with a transition to a higher band. While no such transitions exists for the case of $\tilde{m}=0$ the band difference of the two lowest bands intersects with the excitation from the second to the fourth Bloch band such that the decay process is possible at these momenta. However, the presence of this decay mechanism is contradicted by the observations shown in Fig.\,\ref{fig:08}: No increase of higher-band populations could be observed for any measurement series and the combined atom number in the first and second Brillouin zone remains constant for the measurements of the $\left|1,-1\right\rangle$ as depicted in Fig.\,\hyperlink{fig:09ht}{\ref{fig:09}(c)}.

Here, the described models of band decay in optical lattices fail to explain our observations. However, as an important aspect, the additional influence of the external trapping potential has to be taken into account as it induces a momentum oscillation of the atoms in the excited band. \cite{Heinze2013}. For an oscillation through the minimal band gap atoms can, thus, undergo a Landau-Zener-St{\"u}ckelberg transition into the lowest band. Yet, simulations of such transitions with the harmonic trapping frequencies present in the system yield timescales of such decay processes that are at least an order of magnitude larger as the observed lifetimes. We therefore tend to exclude this scenario as an explanation for the striking difference in the observed lifetimes of atoms in the excited band for opened and closed Dirac cones.

\section{Conclusion \& outlook}\label{sec4}
In conclusion, we have successfully demonstrated a novel approach to continuously alter both the real-space lattice potential as well as the energy band structure of a state-dependent \textit{artificial graphene} lattice. Moreover, by breaking the inversion symmetry of the lattice, an energy gap at the Dirac points can be opened and adjusted continuously. As this experimental technique solely relies on a change of the orientation of the systems quantization field it opens up new possibilities in both the static as well as the dynamic engineering of state-dependent optical potentials in a clean and easily controllable way.

In proof-of-principle experiments, the applicability of the implemented technique could be verified by an \textit{in situ} probing of the state-dependent potential with band-selective microwave spectroscopy. The resulting spectra are in good agreement with numerical calculations of transition frequencies. Furthermore, interaction effects could be described by means of an effective potential in the mean-field regime. Extensive \textit{ab initio} simulations of a microwave spectrum over a large frequency range yield an excellent agreement with the obtained data by including the in-trap density distribution of atoms.

The study of decay mechanisms from the first excited Bloch band revealed a striking influence of the lifetime in the excited Band with respect to the presence of Dirac points. However, further systematic investigations are necessary in order to solve open questions regarding the identification of decay mechanisms.

Beyond the discussed experiments, the newly implemented technique offers many other possible applications. For example, the ability to quickly rotate the quantization axis could allow for excitation schemes similar to Ref.\,\cite{Wirth:ohKKvtz7} and \cite{Oelschlaeger:2011da}, where the sudden change of the anisotropy of sublattices enabled the coherent population of higher Bloch bands. Moreover, the behavior of mixtures of different hyperfine states in a honeycomb lattice can be influenced profoundly by its state-dependency. For example, repulsive interactions between a mixture of two hyperfine states that are trapped at different basis sites leads to a strong \textit{tunneling blockade} in both triangular sublattices \cite{Soltan-Panahi:2011ey}. In addition, such binary spin mixtures can give rise to a new quantum phase in the superfluid regime. It is characterized by a phase twist of the complex superfluid order parameter and results in a symmetry breaking in momentum space \cite{Soltan-Panahi:2011,Luehmann2014hex,Jurgensen:2015}.

In a combination with time-reversal symmetry breaking induced by periodic lattice driving as presented in \cite{Struck:2012gc,Struck:2013ar} the inversion symmetry breaking due to the rotation of the quantization axis could be utilized to investigate topological phase transitions in the framework of the Haldane model \cite{Jotzu2014} or to engineer non-Abelian gauge fields \cite{Hauke:2012dh}. Moreover, the rotation itself can also be utilized for periodic driving schemes in analogy to the spatial translation discussed in the previous chapters. Here, a modulation of the sublattice energy offset may give rise to exotic dynamics in quantum spin-mixtures.

We thank A.~Hemmerich and M.~Lewenstein for stimulating discussions. This work was supported by the Deutsche Forschungsgemeinschaft under grant SFB 925 and the European Community’s Seventh Framework Programme (FP7/2007-2013) under Grant Agreement No. 323714 (EQuaM).


\begin{thebibliography}{31}%
\makeatletter
\providecommand \@ifxundefined [1]{%
 \@ifx{#1\undefined}
}%
\providecommand \@ifnum [1]{%
 \ifnum #1\expandafter \@firstoftwo
 \else \expandafter \@secondoftwo
 \fi
}%
\providecommand \@ifx [1]{%
 \ifx #1\expandafter \@firstoftwo
 \else \expandafter \@secondoftwo
 \fi
}%
\providecommand \natexlab [1]{#1}%
\providecommand \enquote  [1]{``#1''}%
\providecommand \bibnamefont  [1]{#1}%
\providecommand \bibfnamefont [1]{#1}%
\providecommand \citenamefont [1]{#1}%
\providecommand \href@noop [0]{\@secondoftwo}%
\providecommand \href [0]{\begingroup \@sanitize@url \@href}%
\providecommand \@href[1]{\@@startlink{#1}\@@href}%
\providecommand \@@href[1]{\endgroup#1\@@endlink}%
\providecommand \@sanitize@url [0]{\catcode `\\12\catcode `\$12\catcode
  `\&12\catcode `\#12\catcode `\^12\catcode `\_12\catcode `\%12\relax}%
\providecommand \@@startlink[1]{}%
\providecommand \@@endlink[0]{}%
\providecommand \url  [0]{\begingroup\@sanitize@url \@url }%
\providecommand \@url [1]{\endgroup\@href {#1}{\urlprefix }}%
\providecommand \urlprefix  [0]{URL }%
\providecommand \Eprint [0]{\href }%
\providecommand \doibase [0]{http://dx.doi.org/}%
\providecommand \selectlanguage [0]{\@gobble}%
\providecommand \bibinfo  [0]{\@secondoftwo}%
\providecommand \bibfield  [0]{\@secondoftwo}%
\providecommand \translation [1]{[#1]}%
\providecommand \BibitemOpen [0]{}%
\providecommand \bibitemStop [0]{}%
\providecommand \bibitemNoStop [0]{.\EOS\space}%
\providecommand \EOS [0]{\spacefactor3000\relax}%
\providecommand \BibitemShut  [1]{\csname bibitem#1\endcsname}%
\let\auto@bib@innerbib\@empty
\bibitem [{\citenamefont {Novoselov}\ \emph {et~al.}(2005)\citenamefont
  {Novoselov}, \citenamefont {Geim}, \citenamefont {Morozov}, \citenamefont
  {Jiang}, \citenamefont {Katsnelson}, \citenamefont {Grigorieva},
  \citenamefont {Dubonos},\ and\ \citenamefont {Firsov}}]{Novoselov2005}%
  \BibitemOpen
  \bibfield  {author} {\bibinfo {author} {\bibfnamefont {K.~S.}\ \bibnamefont
  {Novoselov}}, \bibinfo {author} {\bibfnamefont {A.~K.}\ \bibnamefont {Geim}},
  \bibinfo {author} {\bibfnamefont {S.~V.}\ \bibnamefont {Morozov}}, \bibinfo
  {author} {\bibfnamefont {D.}~\bibnamefont {Jiang}}, \bibinfo {author}
  {\bibfnamefont {M.~I.}\ \bibnamefont {Katsnelson}}, \bibinfo {author}
  {\bibfnamefont {I.~V.}\ \bibnamefont {Grigorieva}}, \bibinfo {author}
  {\bibfnamefont {S.~V.}\ \bibnamefont {Dubonos}}, \ and\ \bibinfo {author}
  {\bibfnamefont {A.~A.}\ \bibnamefont {Firsov}},\ }\href {\doibase
  10.1038/nature04233} {\bibfield  {journal} {\bibinfo  {journal} {Nature
  (London)}\ }\textbf {\bibinfo {volume} {438}},\ \bibinfo {pages} {197}
  (\bibinfo {year} {2005})}\BibitemShut {NoStop}%
\bibitem [{\citenamefont {Geim}\ and\ \citenamefont
  {Novoselov}(2007)}]{Geim2007}%
  \BibitemOpen
  \bibfield  {author} {\bibinfo {author} {\bibfnamefont {A.~K.}\ \bibnamefont
  {Geim}}\ and\ \bibinfo {author} {\bibfnamefont {K.~S.}\ \bibnamefont
  {Novoselov}},\ }\href {\doibase 10.1038/nmat1849} {\bibfield  {journal}
  {\bibinfo  {journal} {Nature Mat.}\ }\textbf {\bibinfo {volume} {6}},\
  \bibinfo {pages} {183} (\bibinfo {year} {2007})}\BibitemShut {NoStop}%
\bibitem [{\citenamefont {Geim}(2009)}]{Geim2009}%
  \BibitemOpen
  \bibfield  {author} {\bibinfo {author} {\bibfnamefont {A.~K.}\ \bibnamefont
  {Geim}},\ }\href {\doibase 10.1126/science.1158877} {\bibfield  {journal}
  {\bibinfo  {journal} {Science}\ }\textbf {\bibinfo {volume} {324}},\ \bibinfo
  {pages} {1530} (\bibinfo {year} {2009})}\BibitemShut {NoStop}%
\bibitem [{\citenamefont {Soltan-Panahi}\ \emph
  {et~al.}(2011{\natexlab{a}})\citenamefont {Soltan-Panahi}, \citenamefont
  {Struck}, \citenamefont {Hauke}, \citenamefont {Bick}, \citenamefont
  {Plenkers}, \citenamefont {Meineke}, \citenamefont {Becker}, \citenamefont
  {Windpassinger}, \citenamefont {Lewenstein},\ and\ \citenamefont
  {Sengstock}}]{Soltan-Panahi:2011ey}%
  \BibitemOpen
  \bibfield  {author} {\bibinfo {author} {\bibfnamefont {P.}~\bibnamefont
  {Soltan-Panahi}}, \bibinfo {author} {\bibfnamefont {J.}~\bibnamefont
  {Struck}}, \bibinfo {author} {\bibfnamefont {P.}~\bibnamefont {Hauke}},
  \bibinfo {author} {\bibfnamefont {A.}~\bibnamefont {Bick}}, \bibinfo {author}
  {\bibfnamefont {W.}~\bibnamefont {Plenkers}}, \bibinfo {author}
  {\bibfnamefont {G.}~\bibnamefont {Meineke}}, \bibinfo {author} {\bibfnamefont
  {C.}~\bibnamefont {Becker}}, \bibinfo {author} {\bibfnamefont
  {P.}~\bibnamefont {Windpassinger}}, \bibinfo {author} {\bibfnamefont
  {M.}~\bibnamefont {Lewenstein}}, \ and\ \bibinfo {author} {\bibfnamefont
  {K.}~\bibnamefont {Sengstock}},\ }\href {\doibase 10.1038/nphys1916}
  {\bibfield  {journal} {\bibinfo  {journal} {Nature Phys.}\ }\textbf {\bibinfo
  {volume} {7}},\ \bibinfo {pages} {434} (\bibinfo {year}
  {2011}{\natexlab{a}})}\BibitemShut {NoStop}%
\bibitem [{\citenamefont {Soltan-Panahi}\ \emph
  {et~al.}(2011{\natexlab{b}})\citenamefont {Soltan-Panahi}, \citenamefont
  {L{\"u}hmann}, \citenamefont {Struck}, \citenamefont {Windpassinger},\ and\
  \citenamefont {Sengstock}}]{Soltan-Panahi:2011}%
  \BibitemOpen
  \bibfield  {author} {\bibinfo {author} {\bibfnamefont {P.}~\bibnamefont
  {Soltan-Panahi}}, \bibinfo {author} {\bibfnamefont {D.-S.}\ \bibnamefont
  {L{\"u}hmann}}, \bibinfo {author} {\bibfnamefont {J.}~\bibnamefont {Struck}},
  \bibinfo {author} {\bibfnamefont {P.}~\bibnamefont {Windpassinger}}, \ and\
  \bibinfo {author} {\bibfnamefont {K.}~\bibnamefont {Sengstock}},\ }\href
  {\doibase 10.1038/nphys2128} {\bibfield  {journal} {\bibinfo  {journal}
  {Nature Phys.}\ }\textbf {\bibinfo {volume} {8}},\ \bibinfo {pages} {71}
  (\bibinfo {year} {2011}{\natexlab{b}})}\BibitemShut {NoStop}%
\bibitem [{\citenamefont {Tarruell}\ \emph {et~al.}(2012)\citenamefont
  {Tarruell}, \citenamefont {Greif}, \citenamefont {Uehlinger}, \citenamefont
  {Jotzu},\ and\ \citenamefont {Esslinger}}]{Tarruell:2012db}%
  \BibitemOpen
  \bibfield  {author} {\bibinfo {author} {\bibfnamefont {L.}~\bibnamefont
  {Tarruell}}, \bibinfo {author} {\bibfnamefont {D.}~\bibnamefont {Greif}},
  \bibinfo {author} {\bibfnamefont {T.}~\bibnamefont {Uehlinger}}, \bibinfo
  {author} {\bibfnamefont {G.}~\bibnamefont {Jotzu}}, \ and\ \bibinfo {author}
  {\bibfnamefont {T.}~\bibnamefont {Esslinger}},\ }\href {\doibase
  10.1038/nature10871} {\bibfield  {journal} {\bibinfo  {journal} {Nature
  (London)}\ }\textbf {\bibinfo {volume} {483}},\ \bibinfo {pages} {302}
  (\bibinfo {year} {2012})}\BibitemShut {NoStop}%
\bibitem [{\citenamefont {Uehlinger}\ \emph {et~al.}(2013)\citenamefont
  {Uehlinger}, \citenamefont {Jotzu}, \citenamefont {Messer}, \citenamefont
  {Greif}, \citenamefont {Hofstetter}, \citenamefont {Bissbort},\ and\
  \citenamefont {Esslinger}}]{Uehlinger2013}%
  \BibitemOpen
  \bibfield  {author} {\bibinfo {author} {\bibfnamefont {T.}~\bibnamefont
  {Uehlinger}}, \bibinfo {author} {\bibfnamefont {G.}~\bibnamefont {Jotzu}},
  \bibinfo {author} {\bibfnamefont {M.}~\bibnamefont {Messer}}, \bibinfo
  {author} {\bibfnamefont {D.}~\bibnamefont {Greif}}, \bibinfo {author}
  {\bibfnamefont {W.}~\bibnamefont {Hofstetter}}, \bibinfo {author}
  {\bibfnamefont {U.}~\bibnamefont {Bissbort}}, \ and\ \bibinfo {author}
  {\bibfnamefont {T.}~\bibnamefont {Esslinger}},\ }\href {\doibase
  10.1103/PhysRevLett.111.185307} {\bibfield  {journal} {\bibinfo  {journal}
  {Phys. Rev. Lett.}\ }\textbf {\bibinfo {volume} {111}},\ \bibinfo {pages}
  {185307} (\bibinfo {year} {2013})}\BibitemShut {NoStop}%
\bibitem [{\citenamefont {Polini}\ \emph {et~al.}(2013)\citenamefont {Polini},
  \citenamefont {Guinea}, \citenamefont {Lewenstein}, \citenamefont
  {Manoharanand},\ and\ \citenamefont {Pellegrini}}]{Polini2013}%
  \BibitemOpen
  \bibfield  {author} {\bibinfo {author} {\bibfnamefont {M.}~\bibnamefont
  {Polini}}, \bibinfo {author} {\bibfnamefont {F.}~\bibnamefont {Guinea}},
  \bibinfo {author} {\bibfnamefont {M.}~\bibnamefont {Lewenstein}}, \bibinfo
  {author} {\bibfnamefont {H.~C.}\ \bibnamefont {Manoharanand}}, \ and\
  \bibinfo {author} {\bibfnamefont {V.}~\bibnamefont {Pellegrini}},\ }\href
  {http://www.nature.com/nnano/journal/v8/n9/full/nnano.2013.161.html}
  {\bibfield  {journal} {\bibinfo  {journal} {Nature Nanotech.}\ }\textbf
  {\bibinfo {volume} {8}},\ \bibinfo {pages} {625} (\bibinfo {year}
  {2013})}\BibitemShut {NoStop}%
\bibitem [{\citenamefont {{Jotzu}}\ \emph {et~al.}(2014)\citenamefont
  {{Jotzu}}, \citenamefont {{Messer}}, \citenamefont {{Desbuquois}},
  \citenamefont {{Lebrat}}, \citenamefont {{Uehlinger}}, \citenamefont
  {{Greif}},\ and\ \citenamefont {{Esslinger}}}]{Jotzu2014}%
  \BibitemOpen
  \bibfield  {author} {\bibinfo {author} {\bibfnamefont {G.}~\bibnamefont
  {{Jotzu}}}, \bibinfo {author} {\bibfnamefont {M.}~\bibnamefont {{Messer}}},
  \bibinfo {author} {\bibfnamefont {R.}~\bibnamefont {{Desbuquois}}}, \bibinfo
  {author} {\bibfnamefont {M.}~\bibnamefont {{Lebrat}}}, \bibinfo {author}
  {\bibfnamefont {T.}~\bibnamefont {{Uehlinger}}}, \bibinfo {author}
  {\bibfnamefont {D.}~\bibnamefont {{Greif}}}, \ and\ \bibinfo {author}
  {\bibfnamefont {T.}~\bibnamefont {{Esslinger}}},\ }\href
  {http://www.nature.com/nature/journal/v515/n7526/full/nature13915.html}
  {\bibfield  {journal} {\bibinfo  {journal} {Nature (London)}\ }\textbf
  {\bibinfo {volume} {515}},\ \bibinfo {pages} {237} (\bibinfo {year}
  {2014})}\BibitemShut {NoStop}%
\bibitem [{\citenamefont {Duca}\ \emph {et~al.}(2015)\citenamefont {Duca},
  \citenamefont {Li}, \citenamefont {Reitter}, \citenamefont {Bloch},
  \citenamefont {Schleier-Smith},\ and\ \citenamefont {Schneider}}]{Duca2015}%
  \BibitemOpen
  \bibfield  {author} {\bibinfo {author} {\bibfnamefont {L.}~\bibnamefont
  {Duca}}, \bibinfo {author} {\bibfnamefont {T.}~\bibnamefont {Li}}, \bibinfo
  {author} {\bibfnamefont {M.}~\bibnamefont {Reitter}}, \bibinfo {author}
  {\bibfnamefont {I.}~\bibnamefont {Bloch}}, \bibinfo {author} {\bibfnamefont
  {M.}~\bibnamefont {Schleier-Smith}}, \ and\ \bibinfo {author} {\bibfnamefont
  {U.}~\bibnamefont {Schneider}},\ }\href {\doibase 10.1126/science.1259052}
  {\bibfield  {journal} {\bibinfo  {journal} {Science}\ }\textbf {\bibinfo
  {volume} {347}},\ \bibinfo {pages} {288} (\bibinfo {year}
  {2015})}\BibitemShut {NoStop}%
\bibitem [{\citenamefont {Messer}\ \emph {et~al.}(2015)\citenamefont {Messer},
  \citenamefont {Desbuquois}, \citenamefont {Uehlinger}, \citenamefont {Jotzu},
  \citenamefont {Huber}, \citenamefont {Greif},\ and\ \citenamefont
  {Esslinger}}]{Messer2015}%
  \BibitemOpen
  \bibfield  {author} {\bibinfo {author} {\bibfnamefont {M.}~\bibnamefont
  {Messer}}, \bibinfo {author} {\bibfnamefont {R.}~\bibnamefont {Desbuquois}},
  \bibinfo {author} {\bibfnamefont {T.}~\bibnamefont {Uehlinger}}, \bibinfo
  {author} {\bibfnamefont {G.}~\bibnamefont {Jotzu}}, \bibinfo {author}
  {\bibfnamefont {S.}~\bibnamefont {Huber}}, \bibinfo {author} {\bibfnamefont
  {D.}~\bibnamefont {Greif}}, \ and\ \bibinfo {author} {\bibfnamefont
  {T.}~\bibnamefont {Esslinger}},\ }\href {\doibase
  10.1103/PhysRevLett.115.115303} {\bibfield  {journal} {\bibinfo  {journal}
  {Phys. Rev. Lett.}\ }\textbf {\bibinfo {volume} {115}},\ \bibinfo {pages}
  {115303} (\bibinfo {year} {2015})}\BibitemShut {NoStop}%
\bibitem [{\citenamefont {Li}\ \emph {et~al.}(2015)\citenamefont {Li},
  \citenamefont {Duca}, \citenamefont {Reitter}, \citenamefont {Grusdt},
  \citenamefont {Demler}, \citenamefont {Endres}, \citenamefont
  {Schleier-Smith}, \citenamefont {Bloch},\ and\ \citenamefont
  {Schneider}}]{Li2015}%
  \BibitemOpen
  \bibfield  {author} {\bibinfo {author} {\bibfnamefont {T.}~\bibnamefont
  {Li}}, \bibinfo {author} {\bibfnamefont {L.}~\bibnamefont {Duca}}, \bibinfo
  {author} {\bibfnamefont {M.}~\bibnamefont {Reitter}}, \bibinfo {author}
  {\bibfnamefont {F.}~\bibnamefont {Grusdt}}, \bibinfo {author} {\bibfnamefont
  {E.}~\bibnamefont {Demler}}, \bibinfo {author} {\bibfnamefont
  {M.}~\bibnamefont {Endres}}, \bibinfo {author} {\bibfnamefont
  {M.}~\bibnamefont {Schleier-Smith}}, \bibinfo {author} {\bibfnamefont
  {I.}~\bibnamefont {Bloch}}, \ and\ \bibinfo {author} {\bibfnamefont
  {U.}~\bibnamefont {Schneider}},\ }\href {http://arxiv.org/abs/1509.02185}
  {\bibfield  {journal} {\bibinfo  {journal} {arXiv:1509.02185}\ } (\bibinfo
  {year} {2015})}\BibitemShut {NoStop}%
\bibitem [{\citenamefont {Fl\"{a}schner}\ \emph {et~al.}(2015)\citenamefont
  {Fl\"{a}schner}, \citenamefont {Rem}, \citenamefont {Tarnowski},
  \citenamefont {Vogel}, \citenamefont {L\"{u}hmann}, \citenamefont
  {Sengstock},\ and\ \citenamefont {Weitenberg}}]{Flaeschner2015}%
  \BibitemOpen
  \bibfield  {author} {\bibinfo {author} {\bibfnamefont {N.}~\bibnamefont
  {Fl\"{a}schner}}, \bibinfo {author} {\bibfnamefont {B.~S.}\ \bibnamefont
  {Rem}}, \bibinfo {author} {\bibfnamefont {M.}~\bibnamefont {Tarnowski}},
  \bibinfo {author} {\bibfnamefont {D.}~\bibnamefont {Vogel}}, \bibinfo
  {author} {\bibfnamefont {D.-S.}\ \bibnamefont {L\"{u}hmann}}, \bibinfo
  {author} {\bibfnamefont {K.}~\bibnamefont {Sengstock}}, \ and\ \bibinfo
  {author} {\bibfnamefont {C.}~\bibnamefont {Weitenberg}},\ }\href
  {http://arxiv.org/abs/1509.05763} {\bibfield  {journal} {\bibinfo  {journal}
  {arXiv:1509.05763}\ } (\bibinfo {year} {2015})}\BibitemShut {NoStop}%
\bibitem [{\citenamefont {M\"uller}\ \emph {et~al.}(2007)\citenamefont
  {M\"uller}, \citenamefont {F\"olling}, \citenamefont {Widera},\ and\
  \citenamefont {Bloch}}]{Mueller2007}%
  \BibitemOpen
  \bibfield  {author} {\bibinfo {author} {\bibfnamefont {T.}~\bibnamefont
  {M\"uller}}, \bibinfo {author} {\bibfnamefont {S.}~\bibnamefont {F\"olling}},
  \bibinfo {author} {\bibfnamefont {A.}~\bibnamefont {Widera}}, \ and\ \bibinfo
  {author} {\bibfnamefont {I.}~\bibnamefont {Bloch}},\ }\href {\doibase
  10.1103/PhysRevLett.99.200405} {\bibfield  {journal} {\bibinfo  {journal}
  {Phys. Rev. Lett.}\ }\textbf {\bibinfo {volume} {99}},\ \bibinfo {pages}
  {200405} (\bibinfo {year} {2007})}\BibitemShut {NoStop}%
\bibitem [{\citenamefont {Wirth}\ \emph {et~al.}(2011)\citenamefont {Wirth},
  \citenamefont {{\"O}lschl{\"a}ger},\ and\ \citenamefont
  {Hemmerich}}]{Wirth:ohKKvtz7}%
  \BibitemOpen
  \bibfield  {author} {\bibinfo {author} {\bibfnamefont {G.}~\bibnamefont
  {Wirth}}, \bibinfo {author} {\bibfnamefont {M.}~\bibnamefont
  {{\"O}lschl{\"a}ger}}, \ and\ \bibinfo {author} {\bibfnamefont
  {A.}~\bibnamefont {Hemmerich}},\ }\href
  {http://www.nature.com/nphys/journal/v7/n2/full/nphys1857.html} {\bibfield
  {journal} {\bibinfo  {journal} {Nature Phys.}\ }\textbf {\bibinfo {volume}
  {7}},\ \bibinfo {pages} {147} (\bibinfo {year} {2011})}\BibitemShut {NoStop}%
\bibitem [{\citenamefont {{\"O}lschl{\"a}ger}\ \emph
  {et~al.}(2011)\citenamefont {{\"O}lschl{\"a}ger}, \citenamefont {Wirth},\
  and\ \citenamefont {Hemmerich}}]{Oelschlaeger:2011da}%
  \BibitemOpen
  \bibfield  {author} {\bibinfo {author} {\bibfnamefont {M.}~\bibnamefont
  {{\"O}lschl{\"a}ger}}, \bibinfo {author} {\bibfnamefont {G.}~\bibnamefont
  {Wirth}}, \ and\ \bibinfo {author} {\bibfnamefont {A.}~\bibnamefont
  {Hemmerich}},\ }\href {\doibase 10.1103/PhysRevLett.106.015302} {\bibfield
  {journal} {\bibinfo  {journal} {Phys. Rev. Lett.}\ }\textbf {\bibinfo
  {volume} {106}},\ \bibinfo {pages} {15302} (\bibinfo {year}
  {2011})}\BibitemShut {NoStop}%
\bibitem [{\citenamefont {{\"O}lschl{\"a}ger}\ \emph
  {et~al.}(2013)\citenamefont {{\"O}lschl{\"a}ger}, \citenamefont {Kock},
  \citenamefont {Wirth}, \citenamefont {Ewerbeck}, \citenamefont {Smith},\ and\
  \citenamefont {Hemmerich}}]{Oelschlaeger2013a}%
  \BibitemOpen
  \bibfield  {author} {\bibinfo {author} {\bibfnamefont {M.}~\bibnamefont
  {{\"O}lschl{\"a}ger}}, \bibinfo {author} {\bibfnamefont {T.}~\bibnamefont
  {Kock}}, \bibinfo {author} {\bibfnamefont {G.}~\bibnamefont {Wirth}},
  \bibinfo {author} {\bibfnamefont {A.}~\bibnamefont {Ewerbeck}}, \bibinfo
  {author} {\bibfnamefont {C.~M.}\ \bibnamefont {Smith}}, \ and\ \bibinfo
  {author} {\bibfnamefont {A.}~\bibnamefont {Hemmerich}},\ }\href
  {http://stacks.iop.org/1367-2630/15/i=8/a=083041} {\bibfield  {journal}
  {\bibinfo  {journal} {New J. Phys.}\ }\textbf {\bibinfo {volume} {15}},\
  \bibinfo {pages} {083041} (\bibinfo {year} {2013})}\BibitemShut {NoStop}%
\bibitem [{\citenamefont {{\"O}lschl{\"a}ger}\ \emph
  {et~al.}(2012)\citenamefont {{\"O}lschl{\"a}ger}, \citenamefont {Wirth},
  \citenamefont {Kock},\ and\ \citenamefont {Hemmerich}}]{Oelschlaeger:2012fw}%
  \BibitemOpen
  \bibfield  {author} {\bibinfo {author} {\bibfnamefont {M.}~\bibnamefont
  {{\"O}lschl{\"a}ger}}, \bibinfo {author} {\bibfnamefont {G.}~\bibnamefont
  {Wirth}}, \bibinfo {author} {\bibfnamefont {T.}~\bibnamefont {Kock}}, \ and\
  \bibinfo {author} {\bibfnamefont {A.}~\bibnamefont {Hemmerich}},\ }\href
  {\doibase 10.1103/PhysRevLett.108.075302} {\bibfield  {journal} {\bibinfo
  {journal} {Phys. Rev. Lett.}\ }\textbf {\bibinfo {volume} {108}},\ \bibinfo
  {pages} {75302} (\bibinfo {year} {2012})}\BibitemShut {NoStop}%
\bibitem [{\citenamefont {K{\"o}hl}\ \emph {et~al.}(2005)\citenamefont
  {K{\"o}hl}, \citenamefont {Moritz}, \citenamefont {St{\"o}ferle},
  \citenamefont {G{\"u}nter},\ and\ \citenamefont {Esslinger}}]{Koehl:2005cf}%
  \BibitemOpen
  \bibfield  {author} {\bibinfo {author} {\bibfnamefont {M.}~\bibnamefont
  {K{\"o}hl}}, \bibinfo {author} {\bibfnamefont {H.}~\bibnamefont {Moritz}},
  \bibinfo {author} {\bibfnamefont {T.}~\bibnamefont {St{\"o}ferle}}, \bibinfo
  {author} {\bibfnamefont {K.}~\bibnamefont {G{\"u}nter}}, \ and\ \bibinfo
  {author} {\bibfnamefont {T.}~\bibnamefont {Esslinger}},\ }\href {\doibase
  10.1103/PhysRevLett.94.080403} {\bibfield  {journal} {\bibinfo  {journal}
  {Phys. Rev. Lett.}\ }\textbf {\bibinfo {volume} {94}},\ \bibinfo {pages}
  {080403} (\bibinfo {year} {2005})}\BibitemShut {NoStop}%
\bibitem [{\citenamefont {Heinze}\ \emph {et~al.}(2011)\citenamefont {Heinze},
  \citenamefont {G{\"o}tze}, \citenamefont {Krauser}, \citenamefont {Hundt},
  \citenamefont {Fl{\"a}schner}, \citenamefont {L{\"u}hmann}, \citenamefont
  {Becker},\ and\ \citenamefont {Sengstock}}]{Heinze:2011hf}%
  \BibitemOpen
  \bibfield  {author} {\bibinfo {author} {\bibfnamefont {J.}~\bibnamefont
  {Heinze}}, \bibinfo {author} {\bibfnamefont {S.}~\bibnamefont {G{\"o}tze}},
  \bibinfo {author} {\bibfnamefont {J.~S.}\ \bibnamefont {Krauser}}, \bibinfo
  {author} {\bibfnamefont {B.}~\bibnamefont {Hundt}}, \bibinfo {author}
  {\bibfnamefont {N.}~\bibnamefont {Fl{\"a}schner}}, \bibinfo {author}
  {\bibfnamefont {D.~S.}\ \bibnamefont {L{\"u}hmann}}, \bibinfo {author}
  {\bibfnamefont {C.}~\bibnamefont {Becker}}, \ and\ \bibinfo {author}
  {\bibfnamefont {K.}~\bibnamefont {Sengstock}},\ }\href {\doibase
  10.1103/PhysRevLett.107.135303} {\bibfield  {journal} {\bibinfo  {journal}
  {Phys. Rev. Lett.}\ }\textbf {\bibinfo {volume} {107}},\ \bibinfo {pages}
  {135303} (\bibinfo {year} {2011})}\BibitemShut {NoStop}%
\bibitem [{\citenamefont {Heinze}\ \emph {et~al.}(2013)\citenamefont {Heinze},
  \citenamefont {Krauser}, \citenamefont {Fl\"aschner}, \citenamefont {Hundt},
  \citenamefont {G\"otze}, \citenamefont {Itin}, \citenamefont {Mathey},
  \citenamefont {Sengstock},\ and\ \citenamefont {Becker}}]{Heinze2013}%
  \BibitemOpen
  \bibfield  {author} {\bibinfo {author} {\bibfnamefont {J.}~\bibnamefont
  {Heinze}}, \bibinfo {author} {\bibfnamefont {J.~S.}\ \bibnamefont {Krauser}},
  \bibinfo {author} {\bibfnamefont {N.}~\bibnamefont {Fl\"aschner}}, \bibinfo
  {author} {\bibfnamefont {B.}~\bibnamefont {Hundt}}, \bibinfo {author}
  {\bibfnamefont {S.}~\bibnamefont {G\"otze}}, \bibinfo {author} {\bibfnamefont
  {A.~P.}\ \bibnamefont {Itin}}, \bibinfo {author} {\bibfnamefont
  {L.}~\bibnamefont {Mathey}}, \bibinfo {author} {\bibfnamefont
  {K.}~\bibnamefont {Sengstock}}, \ and\ \bibinfo {author} {\bibfnamefont
  {C.}~\bibnamefont {Becker}},\ }\href {\doibase
  10.1103/PhysRevLett.110.085302} {\bibfield  {journal} {\bibinfo  {journal}
  {Phys. Rev. Lett.}\ }\textbf {\bibinfo {volume} {110}},\ \bibinfo {pages}
  {085302} (\bibinfo {year} {2013})}\BibitemShut {NoStop}%
\bibitem [{\citenamefont {L\"uhmann}\ \emph {et~al.}(2014)\citenamefont
  {L\"uhmann}, \citenamefont {J\"urgensen}, \citenamefont {Weinberg},
  \citenamefont {Simonet}, \citenamefont {Soltan-Panahi},\ and\ \citenamefont
  {Sengstock}}]{Luehmann2014hex}%
  \BibitemOpen
  \bibfield  {author} {\bibinfo {author} {\bibfnamefont {D.-S.}\ \bibnamefont
  {L\"uhmann}}, \bibinfo {author} {\bibfnamefont {O.}~\bibnamefont
  {J\"urgensen}}, \bibinfo {author} {\bibfnamefont {M.}~\bibnamefont
  {Weinberg}}, \bibinfo {author} {\bibfnamefont {J.}~\bibnamefont {Simonet}},
  \bibinfo {author} {\bibfnamefont {P.}~\bibnamefont {Soltan-Panahi}}, \ and\
  \bibinfo {author} {\bibfnamefont {K.}~\bibnamefont {Sengstock}},\ }\href
  {\doibase 10.1103/PhysRevA.90.013614} {\bibfield  {journal} {\bibinfo
  {journal} {Phys. Rev. A}\ }\textbf {\bibinfo {volume} {90}},\ \bibinfo
  {pages} {013614} (\bibinfo {year} {2014})}\BibitemShut {NoStop}%
\bibitem [{\citenamefont {Grynberg}\ \emph {et~al.}(1993)\citenamefont
  {Grynberg}, \citenamefont {Lounis}, \citenamefont {Verkerk}, \citenamefont
  {Courtois},\ and\ \citenamefont {Salomon}}]{Grynberg:1993bm}%
  \BibitemOpen
  \bibfield  {author} {\bibinfo {author} {\bibfnamefont {G.}~\bibnamefont
  {Grynberg}}, \bibinfo {author} {\bibfnamefont {B.}~\bibnamefont {Lounis}},
  \bibinfo {author} {\bibfnamefont {P.}~\bibnamefont {Verkerk}}, \bibinfo
  {author} {\bibfnamefont {J.~Y.}\ \bibnamefont {Courtois}}, \ and\ \bibinfo
  {author} {\bibfnamefont {C.}~\bibnamefont {Salomon}},\ }\href {\doibase
  10.1103/PhysRevLett.70.2249} {\bibfield  {journal} {\bibinfo  {journal}
  {Phys. Rev. Lett.}\ }\textbf {\bibinfo {volume} {70}},\ \bibinfo {pages}
  {2249} (\bibinfo {year} {1993})}\BibitemShut {NoStop}%
\bibitem [{\citenamefont {Becker}\ \emph {et~al.}(2010)\citenamefont {Becker},
  \citenamefont {Soltan-Panahi}, \citenamefont {Kronj{\"a}ger}, \citenamefont
  {D{\"o}rscher}, \citenamefont {Bongs},\ and\ \citenamefont
  {Sengstock}}]{Becker:2010de}%
  \BibitemOpen
  \bibfield  {author} {\bibinfo {author} {\bibfnamefont {C.}~\bibnamefont
  {Becker}}, \bibinfo {author} {\bibfnamefont {P.}~\bibnamefont
  {Soltan-Panahi}}, \bibinfo {author} {\bibfnamefont {J.}~\bibnamefont
  {Kronj{\"a}ger}}, \bibinfo {author} {\bibfnamefont {S.}~\bibnamefont
  {D{\"o}rscher}}, \bibinfo {author} {\bibfnamefont {K.}~\bibnamefont {Bongs}},
  \ and\ \bibinfo {author} {\bibfnamefont {K.}~\bibnamefont {Sengstock}},\
  }\href {\doibase 10.1088/1367-2630/12/6/065025} {\bibfield  {journal}
  {\bibinfo  {journal} {New J. Phys.}\ }\textbf {\bibinfo {volume} {12}},\
  \bibinfo {pages} {065025} (\bibinfo {year} {2010})}\BibitemShut {NoStop}%
\bibitem [{\citenamefont {Martikainen}(2011)}]{Martikainen2011}%
  \BibitemOpen
  \bibfield  {author} {\bibinfo {author} {\bibfnamefont {J.-P.}\ \bibnamefont
  {Martikainen}},\ }\href {\doibase 10.1103/PhysRevA.83.013610} {\bibfield
  {journal} {\bibinfo  {journal} {Phys. Rev. A}\ }\textbf {\bibinfo {volume}
  {83}},\ \bibinfo {pages} {013610} (\bibinfo {year} {2011})}\BibitemShut
  {NoStop}%
\bibitem [{\citenamefont {B\"{u}cker}\ \emph {et~al.}(2011)\citenamefont
  {B\"{u}cker}, \citenamefont {Grond}, \citenamefont {Manz}, \citenamefont
  {Berrada}, \citenamefont {Betz}, \citenamefont {Koller}, \citenamefont
  {Hohenester}, \citenamefont {Schumm}, \citenamefont {Perrin},\ and\
  \citenamefont {Schmiedmayer}}]{Buecker2011}%
  \BibitemOpen
  \bibfield  {author} {\bibinfo {author} {\bibfnamefont {R.}~\bibnamefont
  {B\"{u}cker}}, \bibinfo {author} {\bibfnamefont {J.}~\bibnamefont {Grond}},
  \bibinfo {author} {\bibfnamefont {S.}~\bibnamefont {Manz}}, \bibinfo {author}
  {\bibfnamefont {T.}~\bibnamefont {Berrada}}, \bibinfo {author} {\bibfnamefont
  {T.}~\bibnamefont {Betz}}, \bibinfo {author} {\bibfnamefont {C.}~\bibnamefont
  {Koller}}, \bibinfo {author} {\bibfnamefont {U.}~\bibnamefont {Hohenester}},
  \bibinfo {author} {\bibfnamefont {T.}~\bibnamefont {Schumm}}, \bibinfo
  {author} {\bibfnamefont {A.}~\bibnamefont {Perrin}}, \ and\ \bibinfo {author}
  {\bibfnamefont {J.}~\bibnamefont {Schmiedmayer}},\ }\href
  {http://www.nature.com/nphys/journal/v7/n8/full/nphys1992.html} {\bibfield
  {journal} {\bibinfo  {journal} {Nature Physics}\ }\textbf {\bibinfo {volume}
  {7}},\ \bibinfo {pages} {608} (\bibinfo {year} {2011})}\BibitemShut {NoStop}%
\bibitem [{\citenamefont {Paul}\ and\ \citenamefont
  {Tiesinga}(2013)}]{Paul2013}%
  \BibitemOpen
  \bibfield  {author} {\bibinfo {author} {\bibfnamefont {S.}~\bibnamefont
  {Paul}}\ and\ \bibinfo {author} {\bibfnamefont {E.}~\bibnamefont
  {Tiesinga}},\ }\href {\doibase 10.1103/PhysRevA.88.033615} {\bibfield
  {journal} {\bibinfo  {journal} {Phys. Rev. A}\ }\textbf {\bibinfo {volume}
  {88}},\ \bibinfo {pages} {033615} (\bibinfo {year} {2013})}\BibitemShut
  {NoStop}%
\bibitem [{\citenamefont {J\"{u}gensen}\ \emph {et~al.}(2015)\citenamefont
  {J\"{u}gensen}, \citenamefont {Sengstock},\ and\ \citenamefont
  {L\"{u}hmann}}]{Jurgensen:2015}%
  \BibitemOpen
  \bibfield  {author} {\bibinfo {author} {\bibfnamefont {O.}~\bibnamefont
  {J\"{u}gensen}}, \bibinfo {author} {\bibfnamefont {K.}~\bibnamefont
  {Sengstock}}, \ and\ \bibinfo {author} {\bibfnamefont {D.-S.}\ \bibnamefont
  {L\"{u}hmann}},\ }\href {\doibase 10.1038/srep12912} {\bibfield  {journal}
  {\bibinfo  {journal} {Sci. Rep.}\ }\textbf {\bibinfo {volume} {5}},\ \bibinfo
  {pages} {12912} (\bibinfo {year} {2015})}\BibitemShut {NoStop}%
\bibitem [{\citenamefont {Struck}\ \emph {et~al.}(2012)\citenamefont {Struck},
  \citenamefont {{\"O}lschl{\"a}ger}, \citenamefont {Weinberg}, \citenamefont
  {Hauke}, \citenamefont {Simonet}, \citenamefont {Eckardt}, \citenamefont
  {Lewenstein}, \citenamefont {Sengstock},\ and\ \citenamefont
  {Windpassinger}}]{Struck:2012gc}%
  \BibitemOpen
  \bibfield  {author} {\bibinfo {author} {\bibfnamefont {J.}~\bibnamefont
  {Struck}}, \bibinfo {author} {\bibfnamefont {C.}~\bibnamefont
  {{\"O}lschl{\"a}ger}}, \bibinfo {author} {\bibfnamefont {M.}~\bibnamefont
  {Weinberg}}, \bibinfo {author} {\bibfnamefont {P.}~\bibnamefont {Hauke}},
  \bibinfo {author} {\bibfnamefont {J.}~\bibnamefont {Simonet}}, \bibinfo
  {author} {\bibfnamefont {A.}~\bibnamefont {Eckardt}}, \bibinfo {author}
  {\bibfnamefont {M.}~\bibnamefont {Lewenstein}}, \bibinfo {author}
  {\bibfnamefont {K.}~\bibnamefont {Sengstock}}, \ and\ \bibinfo {author}
  {\bibfnamefont {P.}~\bibnamefont {Windpassinger}},\ }\href {\doibase
  10.1103/PhysRevLett.108.225304} {\bibfield  {journal} {\bibinfo  {journal}
  {Phys. Rev. Lett.}\ }\textbf {\bibinfo {volume} {108}},\ \bibinfo {pages}
  {225304} (\bibinfo {year} {2012})}\BibitemShut {NoStop}%
\bibitem [{\citenamefont {Struck}\ \emph {et~al.}(2013)\citenamefont {Struck},
  \citenamefont {Weinberg}, \citenamefont {{\"O}lschl{\"a}ger}, \citenamefont
  {Windpassinger}, \citenamefont {Simonet}, \citenamefont {Sengstock},
  \citenamefont {H{\"o}ppner}, \citenamefont {Hauke}, \citenamefont {Eckardt},
  \citenamefont {Lewenstein},\ and\ \citenamefont {Mathey}}]{Struck:2013ar}%
  \BibitemOpen
  \bibfield  {author} {\bibinfo {author} {\bibfnamefont {J.}~\bibnamefont
  {Struck}}, \bibinfo {author} {\bibfnamefont {M.}~\bibnamefont {Weinberg}},
  \bibinfo {author} {\bibfnamefont {C.}~\bibnamefont {{\"O}lschl{\"a}ger}},
  \bibinfo {author} {\bibfnamefont {P.}~\bibnamefont {Windpassinger}}, \bibinfo
  {author} {\bibfnamefont {J.}~\bibnamefont {Simonet}}, \bibinfo {author}
  {\bibfnamefont {K.}~\bibnamefont {Sengstock}}, \bibinfo {author}
  {\bibfnamefont {R.}~\bibnamefont {H{\"o}ppner}}, \bibinfo {author}
  {\bibfnamefont {P.}~\bibnamefont {Hauke}}, \bibinfo {author} {\bibfnamefont
  {A.}~\bibnamefont {Eckardt}}, \bibinfo {author} {\bibfnamefont
  {M.}~\bibnamefont {Lewenstein}}, \ and\ \bibinfo {author} {\bibfnamefont
  {L.}~\bibnamefont {Mathey}},\ }\href {\doibase 10.1038/nphys2750} {\bibfield
  {journal} {\bibinfo  {journal} {Nat. Phys.}\ }\textbf {\bibinfo {volume}
  {9}},\ \bibinfo {pages} {738} (\bibinfo {year} {2013})}\BibitemShut {NoStop}%
\bibitem [{\citenamefont {Hauke}\ \emph {et~al.}(2012)\citenamefont {Hauke},
  \citenamefont {Tieleman}, \citenamefont {Celi}, \citenamefont
  {{\"O}lschl{\"a}ger}, \citenamefont {Simonet}, \citenamefont {Struck},
  \citenamefont {Weinberg}, \citenamefont {Windpassinger}, \citenamefont
  {Sengstock}, \citenamefont {Lewenstein},\ and\ \citenamefont
  {Eckardt}}]{Hauke:2012dh}%
  \BibitemOpen
  \bibfield  {author} {\bibinfo {author} {\bibfnamefont {P.}~\bibnamefont
  {Hauke}}, \bibinfo {author} {\bibfnamefont {O.}~\bibnamefont {Tieleman}},
  \bibinfo {author} {\bibfnamefont {A.}~\bibnamefont {Celi}}, \bibinfo {author}
  {\bibfnamefont {C.}~\bibnamefont {{\"O}lschl{\"a}ger}}, \bibinfo {author}
  {\bibfnamefont {J.}~\bibnamefont {Simonet}}, \bibinfo {author} {\bibfnamefont
  {J.}~\bibnamefont {Struck}}, \bibinfo {author} {\bibfnamefont
  {M.}~\bibnamefont {Weinberg}}, \bibinfo {author} {\bibfnamefont
  {P.}~\bibnamefont {Windpassinger}}, \bibinfo {author} {\bibfnamefont
  {K.}~\bibnamefont {Sengstock}}, \bibinfo {author} {\bibfnamefont
  {M.}~\bibnamefont {Lewenstein}}, \ and\ \bibinfo {author} {\bibfnamefont
  {A.}~\bibnamefont {Eckardt}},\ }\href {\doibase
  10.1103/PhysRevLett.109.145301} {\bibfield  {journal} {\bibinfo  {journal}
  {Phys. Rev. Lett.}\ }\textbf {\bibinfo {volume} {109}},\ \bibinfo {pages}
  {145301} (\bibinfo {year} {2012})}\BibitemShut {NoStop}%
\end{thebibliography}
\end{document}